\documentclass[manuscript]{acmart}

\AtBeginDocument{%
  \providecommand\BibTeX{{%
    \normalfont B\kern-0.5em{\scshape i\kern-0.25em b}\kern-0.8em\TeX}}}

\copyrightyear{2024}
\acmYear{2024}
\setcopyright{rightsretained}
\acmConference[ASSETS '24]{The 26th International ACM SIGACCESS Conference on Computers and Accessibility}{October 27--30, 2024}{St. John's, NL, Canada}
\acmBooktitle{The 26th International ACM SIGACCESS Conference on Computers and Accessibility (ASSETS '24), October 27--30, 2024, St. John's, NL, Canada}
\acmDOI{10.1145/3663548.3675662}
\acmISBN{979-8-4007-0677-6/24/10}

\author{Franklin Mingzhe Li}
\affiliation{%
  \institution{Carnegie Mellon University}
  \city{Pittsburgh}
  \state{PA}
  \country{United States}
}
\email{mingzhe2@cs.cmu.edu}

\author{Ashley Wang}
\affiliation{%
  \institution{Carnegie Mellon University}
  \city{Pittsburgh}
  \state{PA}
  \country{United States}
}
\email{awwang@andrew.cmu.edu}

\author{Patrick Carrington}
\affiliation{%
  \institution{Carnegie Mellon University}
  \city{Pittsburgh}
  \state{PA}
  \country{United States}
}
\email{pcarrington@cmu.edu}

\author{Shaun K. Kane}
\affiliation{%
  \institution{Google Research}
  \city{Boulder}
  \state{Colorado}
  \country{United States}
}
\email{shaunkane@google.com}
    
\begin{document}


\title[Non-Visual Access to Cooking Instruction]{A Recipe for Success? Exploring Strategies for Improving Non-Visual Access to Cooking Instructions}

\renewcommand{\shortauthors}{Li et al.}

\begin{abstract}
Cooking is an essential activity that enhances quality of life by enabling individuals to prepare their own meals. However, cooking often requires multitasking between cooking tasks and following instructions, which can be challenging to cooks with vision impairments if recipes or other instructions are inaccessible. To explore the practices and challenges of recipe access while cooking, we conducted semi-structured interviews with 20 people with vision impairments who have cooking experience and four cooking instructors at a vision rehabilitation center. We also asked participants to edit and give feedback on existing recipes. We revealed unique practices and challenges to accessing recipe information at different cooking stages, such as the heavy burden of hand-washing to interact with recipe readers. We also presented the preferred information representation and structure of recipes. We then highlighted design features of technological supports that could facilitate the development of more accessible kitchen technologies for recipe access. Our work contributes nuanced insights and design guidelines to enhance recipe accessibility for people with vision impairments.

\end{abstract}




\maketitle

\section{Introduction}

Cooking independently is a common activity of daily living across the world. Being able to prepare food is a major contributor to a sense of self-efficacy and quality of life. Cooking often requires multitasking, as cooks must prepare food (often through time-sensitive tasks) while following an overall plan, either memorized or by following a video or recipe. Recipes, whether presented as text or video, contain important procedural and contextual details that are often essential to correctly preparing dishes \cite{chang2018recipescape,min2019survey}. Increasingly, home cooks are looking beyond cookbooks and relying on online cooking resources such as blogs and videos as part of their home cooking toolkit~\cite{cooper2015cooking}.

Cooking can present a variety of accessibility challenges to blind and visually impaired cooks, including inaccessible tools, ingredients, and workspaces~\cite{wang2023practices,li2024contextual,li2021non}. Recipes may be inaccessible when they include instructions that expect the cook to be sighted. For example, a recipe might include an instruction to cook an item until it is a particular color (e.g., cook the chicken until golden brown). The lack of non-visual alternatives and descriptions create obstacles to cooking for people with vision impairments \cite{bilyk2009food,kostyra2017food}. Barriers to cooking one’s own food can lead to over-reliance on takeout or prepared foods, which can lead to poor nutritional outcomes \cite{jones2019analysis}.

While prior research has taken a holistic approach to understanding accessibility challenges in the kitchen (e.g.,~\cite{li2021non}), there is value in directly exploring issues related to recipe accessibility, as changes in how home cooks use recipes may perpetuate or increase inaccessibility. Furthermore, incorporating ``smart'' technologies into the kitchen may present opportunities to improve autonomy for people with disabilities~\cite{blasco2014smart}. It seems likely that these accessibility challenges may be partially addressed through the introduction of technology, such as tactile displays \cite{niran2016access,russomanno2015refreshing}, smart speakers \cite{AmazonAl14:online}, computer vision systems \cite{guo2016vizlens}, crowdsourced human assistants~\cite{bigham2010vizwiz,HomeAira36:online,BeMyEyes54:online}, and AI assistants~\cite{gonzalez2024investigating}; however, identifying the most promising opportunities first requires an exploration of current recipe use by people with vision impairments, specifically exploring how cooks with vision impairments currently use recipes and other information sources, accessibility challenges they encounter while doing so, and their wishes for future technology.


To explore these questions, we conducted semi-structured interviews with 20 people with vision impairments who have experience with cooking, along with interviewing four cooking instructors from a vision rehabilitation center. Our research was guided by the following research questions:
\begin{itemize}
    \item RQ1: What methods, strategies, and challenges do people with vision impairments have when accessing recipe information while cooking?
    \item RQ2: What types of recipe information, descriptions, alternatives, and structures are most helpful during the cooking process?
    \item RQ3: How might improvements in technology improve the accessibility of cooking instructions?
\end{itemize}

By exploring these questions, our work provides insight into how people with vision impairments currently use recipes while cooking (Section \ref{findingRQ1}), introduces the preferred design of recipe content and structure (Section \ref{findingRQ2}), and identifies opportunities to create more accessible home cooking experiences (Section \ref{findingRQ3}).

\section{Related Work}

\subsection{Accessibility of Cooking and Meal Preparation for People with Vision Impairments}

People with vision impairments often choose food options in part to avoid accessibility problems~\cite{bilyk2009food}. Through a survey of over 100 visually impaired people, Jones et al. \cite{jones2019analysis} found a significant correlation between the severity of vision impairment and difficulty in shopping for ingredients and cooking meals. Vision impairments are related to malnourishment and poor quality of life \cite{jones2019analysis}. Through semi-structured interviews with nine visually impaired people in 2009, Bilyk et al. \cite{bilyk2009food} found that people with vision impairments heavily rely on prepared food from outside sources, such as restaurants, mostly in order to avoid cooking. Kostyra et al. \cite{kostyra2017food} found that out of 250 survey respondents, 58\% of participants usually consume ready-to-eat products \cite{kostyra2017food}. 

Other research has identified specific accessibility challenges in the kitchen. Bilyk et al. \cite{bilyk2009food} uncovered various challenges that visually impaired individuals encounter while cooking, including accessing cooking instructions and recipes. Li et al. \cite{li2021non} mentioned that accessing recipes can affect how people with vision impairments learn, prepare, and perform cooking actions, specifically due to inaccessible recipe content \cite{li2021non}. This research extends prior work to focus on accessibility challenges related to accessing and following recipes.

\subsection{Accessible Technology in the Kitchen}

Prior research has explored a variety of approaches to provide non-visual access to kitchen equipment, including using computer vision \cite{guo2016vizlens,fusco2014using,morris2006clearspeech,tekin2011real}, voice interactions \cite{abdolrahmani2018siri,branham2019reading,azenkot2013exploring}, and tactile marking \cite{guo2017facade,he2017tactile} to better support visually impaired individuals interacting with different kitchen appliances. For example, Guo et al. \cite{guo2016vizlens} leveraged computer vision and crowd workers to support people with vision impairments to interact with interfaces of kitchen appliances, such as microwave oven. Beyond making appliance interfaces accessible, prior research has also explored various approaches to support people with vision impairments to interact with devices through different gestural interactions \cite{kane2008slide,azenkot2012passchords,li2017braillesketch,kane2011usable}. However, it is unknown how these technologies can be adopted and designed to better support people with vision impairments interacting with recipe information.

\subsection{Electronic Recipes and Cooking Instructions}
Recipes act as the main source of cooking activities \cite{borghini2015recipe}. Prior research has been conducted in order to explore recipe wording and construction. For example, Tasse and Smith \cite{tasse2008sour} created CURD, a database of recipes annotated using the MILK language, a meaning representation language based on first-order logic, to break recipes and steps into pieces of a specific format and function. Various online services that can recommend a recipe based on a set of ingredients exist, such as the Kroger Chefbot \cite{KrogerCh9:online} and ChatGPT \cite{Introduc4:online}. Prior research further explored algorithms to provide more adaptive recipes based on individual needs \cite{teng2011reciperecommendation}, cooking difficulty \cite{kusu2017calculating}, and comparison of similar recipes \cite{chang2018recipescape}. Teng et al. \cite{teng2011reciperecommendation} measured the role of ingredients in recipes, as well as the relationships between co-occurring ingredients. They constructed networks to capture these relationships and used these networks, as well as user feedback, to create recipe recommendation algorithms. Instead of looking at ingredients as previous studies did, Kusu et al. \cite{kusu2017calculating} calculated the difficulty of recipes based on the cooking activities involved and proposed formulas that related cooking activity difficulty level with recipe skill level \cite{kusu2017calculating}. Given these approaches to extracting recipe information and providing adaptive and customized recipes, it is unknown how these can be leveraged for visually impaired individuals, and more importantly, what are the procedural and contextual information needs by people with vision impairments and why.

Practically, there are different ways of accessing recipe information for cooking, including various online resources to find recipes (e.g., AllRecipes.com \cite{Allrecip74:online}, FoodNetwork.com \cite{RecipesD31:online}), traditional cookbooks \cite{davis2014homemade}, cooking videos on YouTube \cite{51YouTub12:online}, or even smart speakers \cite{AmazonAl14:online}. There are also recipes and personal blogs on which users can post any recipes that they wish to share \cite{BowTiePa2:online}. Given majority of recipes are text-based, prior research has explored systems to support people with vision impairments access text-based content, such as screen readers \cite{rodrigues2015getting,Accessib51:online,leporini2012interacting,Getstart6:online,lazar2007frustrates}, speech synthesizers \cite{blenkhorn1995producing,shimomura2010accessibility,guerreiro2014text}, and Braille displays \cite{niran2016access,russomanno2015refreshing}. On the other hand, some recipes might be stored in tactile, image, or video format. prior research also explored different ways to support visually impaired people access tactile \cite{phutane2022tactile}, video \cite{liu2021makes,wang2021toward,feiz2019towards} and image content \cite{morris2016most,harada2013accessible}. Based on existing accessing methods to different forms of content, it is unknown how do people with vision impairments currently access recipe information, as well as the design features of systems for non-visual access to recipes while cooking.

\begin{table*}[]
\resizebox{\textwidth}{!}{%
\begin{tabular}{p{1.5cm}|p{0.5cm}p{1cm}p{6cm}p{2.5cm}p{2cm}p{2.1cm}}
\hline
\textbf{Participant}     & \textbf{Age} & \textbf{Gender} & \textbf{Vision Self-Description} & \textbf{Years at Current Vision Level}                         & \textbf{Years Cooking} & \textbf{Days per Week Cooking} \\ \hline
\multicolumn{1}{l|}{P1}  & 32           & Female          & Totally Blind  &  21  & 15 & 3           \\
\multicolumn{1}{l|}{P2}  & 61           & Male            & Totally Blind & 22                                             & 53               & 4-5           \\
\multicolumn{1}{l|}{P3}  & 48           & Female          & Totally Blind & 48                                              & 20               & 7                   \\
\multicolumn{1}{l|}{P4}  & 57           & Female          & Totally Blind &  42                                         & 40                & 2               \\
\multicolumn{1}{l|}{P5}  & 47           & Male            & Legally Blind with light perception & 47                                              & 23               & Several       \\
\multicolumn{1}{l|}{P6}  & 38           & Male            & Totally blind in one eye, Legally Blind in other & 5 & 32               & Several       \\
\multicolumn{1}{l|}{P7}  & 44           & Female          & Totally Blind, some light perception in one eye                   & 44    & 20               & 4-5          \\
\multicolumn{1}{l|}{P8}  & 26           & Male            & Legally Blind & 6                    & 20               & 2-3           \\
\multicolumn{1}{l|}{P9}  & 31           & Female          & Totally blind in one eye, Legally Blind in other & 4                                             & 18                & 3             \\
\multicolumn{1}{l|}{P10} & 61           & Female          & Totally Blind & 61                                              & 45                & 7                   \\
\multicolumn{1}{l|}{P11} & 38           & Male            & Totally Blind  & 10                                              & 30                & 7                   \\
\multicolumn{1}{l|}{P12} & 67           & Male            & Legally Blind & 67                                              & 49                & 7                   \\
\multicolumn{1}{l|}{P13} & 36           & Male            & Legally Blind & 5                                              & about 20         & 7                   \\
\multicolumn{1}{l|}{P14} & 30           & Female          & Totally Blind  & 22                                              & 11               & once per month               \\
\multicolumn{1}{l|}{P15} & 70           & Female          & Totally Blind  & 54                                             & 58                & 7                   \\
\multicolumn{1}{l|}{P16} & 24           & Female          & Legally Blind, & 24  & 2                & 3-4           \\
\multicolumn{1}{l|}{P17} & 39           & Female          & Totally Blind  & 13                                 & 26                & 7                   \\
\multicolumn{1}{l|}{P18} & 48           & Female          & Totally Blind with very little light perception & 48                                      & 40               & 3-4           \\
\multicolumn{1}{l|}{P19} & 46           & Male            & Totally Blind  & 46        & 29               & 7                   \\
\multicolumn{1}{l|}{P20} & 24           & Male            & Legally Blind & 12                                             & 14                & 3             \\ \hline
\end{tabular}%
}
\caption{Home cooks with vision impairments who participated in our study}
\label{table:participants}
\end{table*}

\section{Method: Interviews with Home Cooks \& Adaptive Cooking Instructors}
\label{Method: Interview Study with Blind Cooks}

We conducted semi-structured interviews with blind people who have experience in cooking, and with adaptive cooking instructors from a vision rehabilitation center. Participants were compensated with a \$20 gift card. The recruitment and study procedures were approved by our Institutional Review Board (IRB). The semi-structured interviews were conducted through Zoom \cite{VideoCon42:online}, and all interviews were audio-recorded and transcribed, verbatim.

\subsection{Interview with Blind Cooks}
\subsubsection{Participants}
We recruited 20 blind participants (Table \ref{table:participants}) through social media (Twitter and Reddit), and mailing lists maintained by blindness organizations. All participants were 18 years or older, legally or totally blind, experienced in cooking, and able to communicate in English. Table \ref{table:participants} presents participants' demographic information.

\subsubsection{Pre-interview Survey}
We asked participants to complete a pre-study questionnaire to confirm that they met the inclusion criteria and asked for demographic information, cooking frequency, recipe sources they use, and three dishes they would like to learn. The questionnaire was administered using a Google Form \cite{GoogleFo84:online} and is included in the supplementary material. 


\subsubsection{Home Cook Interview}
Each interview session took approximately 90 minutes and covered the following topics:
\begin{enumerate}
    \item \textbf{{Demographics and Background [$\sim$ 5 Minutes]:}}
participants' age, gender, descriptions of vision level, and cooking experience. (see Table \ref{table:participants});
    \item \textbf{Current Cooking Practices [$\sim$ 20 Minutes]:}
participants' current ways of accessing recipes and cooking information, such as what technologies they use to search for cooking-related information, how they gather information about physical objects while cooking, and how they adapt information resources to make them accessible;
    \item \textbf{Recipe Review and Editing [$\sim$ 30 Minutes]:} participants edited recipes as described in the next section (Section \ref{Recipe Review and Collaborative Editing});
\label{information needs methods}
    \item \textbf{Current \& Desired Technology Use [$\sim$ 30 Minutes]}
participants' experiences with different technologies and services to access recipe information while cooking.
\end{enumerate}

\subsubsection{Recipe Review and Collaborative Editing}
\label{Recipe Review and Collaborative Editing}
As part of the interview, we asked participants to read and reflect on recipes that they were unfamiliar with, but that reflected dishes they were interested in learning to cook. We used this activity as a probe to get concrete feedback about recipe content and structure, while choosing recipes that matched participants' interest and abilities.

During the pre-interview survey, we asked participants to name three dishes that they would be interested in learning to cook. Before the interview session, we located recipes for each of the dishes through recommended recipe sources by participants, and edited them into a simplified text format. During the session, we walked through the recipes with participants and collected their feedback about each step. This process also informed our follow-up questions during the rest of the interview.



\subsection{Interview with Adaptive Cooking Instructors}
\subsubsection{Participants}
To complement feedback from cooks themselves, we interviewed four instructors (I1 - I4) who teach blind cooking courses at a vision rehabilitation center (SAVVI, located in Arizona, USA); this included totally blind, legally blind, and sighted instructors. Their demographic information is shown in Table \ref{table:instructor}.



\begin{table*}[]
\resizebox{\columnwidth}{!}{%
\begin{tabular}{l|rllll}
\textbf{Participant} & \multicolumn{1}{l}{\textbf{Age}} & \textbf{Gender} & \textbf{Vision Description}               & \textbf{Years at Current Vision Level} & \textbf{Instruction Experience} \\ \hline
I1                   & 58                               & Female          & Legally Blind                             & 4                                      & 9 months                        \\
I2                   & 24                               & Female          & Sighted                                   & 24                                     & 10 months                       \\
I3                   & 58                               & Male            & Legally Blind, can only see high contrast & 3                                      & 4 months                        \\
I4                   & 65                               & Female          & Totally Blind                             & 35                                     & 30 years                        \\ \hline
\end{tabular}%
}
\caption{Instructor Demographic Information}
\label{table:instructor}
\end{table*}

\subsubsection{Adaptive Cooking Instructor Interview}
Each interview session took approximately 75 minutes and covered the following topics:

\begin{itemize}
    \item \textbf{Demographics and Background [$\sim$ 5 Minutes]:} participants' age, gender, vision level, and length of teaching experience; 
    \item \textbf{Experiences in Blind Cooking Instruction [$\sim$ 30 Minutes]:} experiences of instructing blind people in cooking classes, including instructional practices, challenges faced by their students, learning barriers encountered by their students, and preferences for recipe information and instruction; 
    \item \textbf{Perceptions of Recipe Presentation Quality [$\sim$ 20 Minutes]:} perceptions of recipe information sources, quality, and quality of presentation for blind cooks; how information should be presented to blind cooks at different stages of their learning processes and cooking activities;
    \item \textbf{Instruction Information Access Technologies [$\sim$ 20 Minutes]:}
technologies that participants used to teach blind people to cook, and suggestions for how technology could be designed to better support blind cooks.
\end{itemize}

\subsection{Data Analysis}
To analyze the interview data from blind cooks, two researchers performed open coding \cite{charmaz2006constructing} on the same four transcripts to form and reconcile the initial codebook. Those researchers independently coded the remaining transcripts before discussing the codes and resolving conflicts (e.g., missing codes, disagreements). After the two researchers reached a consensus and reconciled the codebook, they performed affinity diagramming \cite{hartson2012ux} using a Miro board \cite{AnOnline70:online} to cluster the codes and identify emergent themes. The interview data from the instructors' interviews were coded separated from the cook interviews but using a similar process of redundant coding for all four transcripts (I1 - I4) by two researchers. This was followed by discussion, conflict resolution, reconciliation, and affinity diagramming to identify emergent themes.

\section{FINDINGS: Current Use of Recipes while Cooking (RQ1)}
\label{findingRQ1}
\begin{figure*}[t]
    \centering
    \includegraphics[width=\columnwidth]{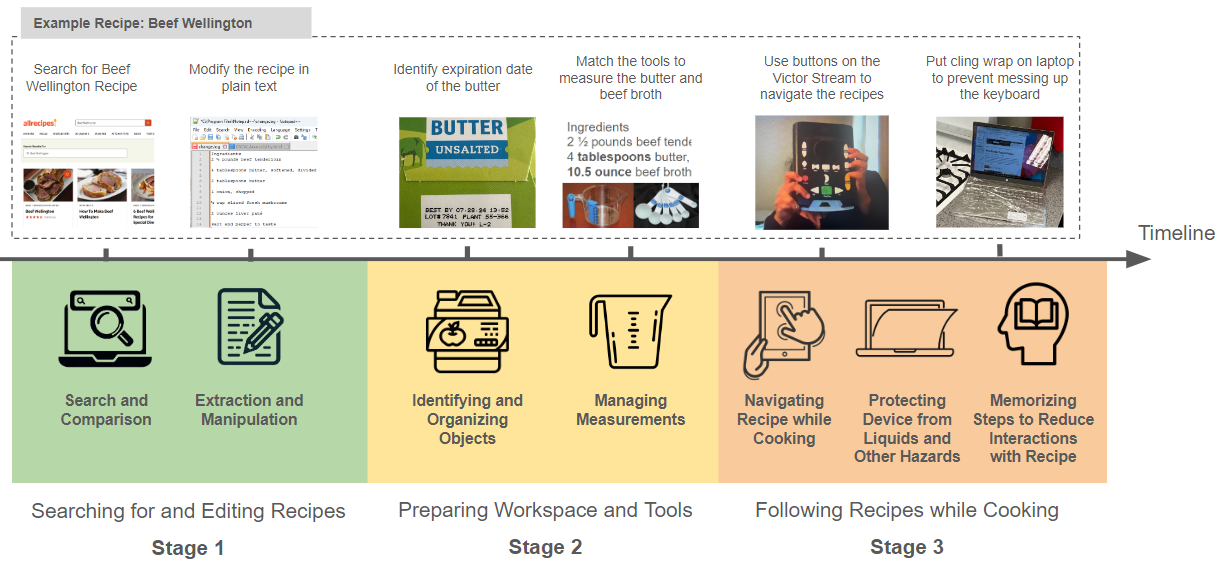}
    \caption{Illustration of cooking phases using example from process to cook Beef Wellington from a Recipe}
    \label{fig:timeline}
    \Description{This figure shows a timeline of the process of leveraging recipes for cooking. Under the timeline axis, there are three stages: 1) stage 1 is searching for and Editing recipes, which include two icons indicating search and comparison, and extraction and manipulation; 2) stage 2 is preparing workspace and tools, which include identifying and organizing objects, and managing measurements; 3) stage 3 is following recipe while cooking, which includes navigating recipe while cooking, protecting device from liquids and other hazards, and memorizing steps to reduce interactions with recipe. Above the timeline axis, there are six steps to using a recipe of beef wellington that include: search for beef wellington recipe, modify the recipe in plain text, identify the expiration date of the butter, match the tools to measure the butter and beef broth, use buttons on the victor stream to navigate the recipe, and put cling wrap on a laptop to prevent messing up the keyboards.}
\end{figure*}

We identified current methods, strategies, and challenges of accessing recipes by cooks with vision impairments. These are loosely organized into three cooking phases (Figure \ref{fig:timeline}): 1) searching for and editing recipes before cooking, 2) preparing their workspace and tools using recipe information, and 3) cooking while following recipe information.

\subsection{Searching for and Editing Recipe Information}
This stage of the process is characterized by actions taken to acquire information needed to perform specific cooking tasks, decide between different recipes, and manipulate the recipes themselves to be used in the kitchen.

\subsubsection{Search and Comparison}
\label{information seeking}
Participants described their ways of searching for recipe information ahead of cooking. All participants mentioned that they use text-based online recipe information sources to learn non-visual cooking skills (e.g., measurement techniques) and new recipes and cuisines. 20 participants described accessing recipes through general purpose recipe websites (e.g., allrecipes.com, foodnetwork.com), 8 mentioned using specialized websites for blind cooks (e.g., visionaware.org, acb.org), and 6 used Facebook groups (e.g., incredible recipes, the country cook). P17 described their experience of leveraging various online information for cooking with relative success:

\begin{quote}
    ``...The ads are really frustrating, but once you can get through those, I've had a lot of success. At Thanksgiving, I always forget how long to do a Turkey or whatever, so I get the weight of my Turkey and I go search and find all these recipes and things like that, and they've been super helpful and I've even gotten some new ideas that I didn't necessarily know about from reading these recipes and things like that. Get a little more adventurous in my cooking. I'm a great basic cook, but trying new ideas and stuff, recipes are great for that...''
\end{quote}

A common challenge was how \textbf{inaccessible and unpredictable visual content in online sources (e.g., ads, commercial videos) diminishes the accessibility of the resource itself}. P6 said:

\begin{quote}
    ``...When navigating the recipe websites, swipe right to scroll through them. Sometimes that would cause problems. If there was a video in the recipe, it would automatically start playing with sound. And even if I were to scroll past it, the video was already playing. Then the sound would just continue...I don't know whether or not the video was still visually on the screen or not. But the audio still [played]. So that was a real hindrance. Especially when you're relying on the audio of TalkBack at the same time, so I can't remember if it was playing over TalkBack, or if TalkBack wasn't working, but I do remember that it basically stopped me in my tracks...''
\end{quote}

Participants often \textbf{search for multiple candidates and compare them based on ingredients, timing, and availability}. P7 explained:

\begin{quote}
    ``...I honestly just hop on Google...I'll look to see how many people have rated it. I'll look through three or four recipes about the same thing, and figure out what sounds good to me, what flavors sound right, what timing sounds right...''
\end{quote}

In addition to accessing recipe information through websites, some participants have tried to use video and audio sources, often with limited success. Twelve participants mentioned that they have accessed recipes through YouTube, but do not rely on it due to \textbf{lack of audio description of actions in YouTube videos, which caused missing information}. Some still use YouTube for ideation and inspiration. Five participants mentioned using audio podcasts for cooking information, but did not generally follow recipes from those podcasts. When we asked participants what devices they used to access this information, most used their phones. Six participants also mentioned that they sometimes used smart speakers to access procedural and contextual information of cooking, but generally used them to answer questions or provide short directions rather than following a full recipe. When asked about challenges related to audio and video cooking information, participants mentioned the difficulty of following recipes with multiple steps (P1, P9), lack of consistency of the recipe sources (P5, P17), and difficulty in extracting recipe information into a format they could follow along with (P5, P12, P19). 

\subsubsection{Extraction and Manipulation}
\label{Instructional information extraction and manipulation}
Once they selected a recipe, our participants applied a variety of strategies to extract and manipulate information to make the content more accessible and easy to navigate. Nine participants described \textbf{saving information either in a Word document or text file}, both for ease of navigation and to avoid ads. P1 commented on how she saved  recipes from Facebook groups: 

\begin{quote}
    ``...So I have joined a bunch of Facebook groups and I'll typically get recipes from those groups, and if I see something I will save it, and typically I'll not look at it at the moment. I'll just sort of save and curate. And then when I have time, I'll go look at the websites and copy everything out into easier-to-read plain text files, so that I don't have to deal with ads and web formatting when I'm trying to cook...''
\end{quote}

In addition to copying and pasting text into documents, participants \textbf{manually typed information from audio or video recipes into a text document}, although doing so required some tedious effort. P16 explained: 

\begin{quote}
    ``...I'll just play it a couple of seconds at a time. If she's listing the ingredients, I'll start it and pause it and then type it into my computer. It's not convenient, but it's nicer than if I miss an ingredient. I'd have to watch the whole video again...''
\end{quote}

After converting the information into text, 13 participants described how they \textbf{edited ingredients and steps}, as well as adding personal notes. Participants usually saved and edited recipes on a desktop or laptop computer, then accessed those recipes on a mobile device while preparing and cooking. Participants often use their screen reader's search function to \textbf{search for headers or specific text, such as an ingredient name or measurement unit}. P1 commented:

\begin{quote}
    ``...I'll have to do a control F and find and I'll search for a word of ingredients or I'll search for the word `cup' because I know that's going to be in there somewhere. That is often why I'll go ahead and copy it into Notepad, because especially on mobile, it's a different experience...I'd rather do it when I'm sitting down and have the time to copy it out...''
\end{quote}

\subsection{Preparing Workspace and Tools}
\label{information identification}
After finding the right recipe and converting it into an accessible format, participants often start the cooking process itself by preparing and laying out tools and ingredients; sometimes this included specific tasks to make objects and ingredients more accessible.

\subsubsection{Identifying and Organizing Objects}
Participants noted that preparing the workspace usually involves \textbf{visually identifying objects and food packaging} such as the name of the object, expiration date, nutritional information, and suggested cooking instructions. Participants use~\textbf{visual recognition apps (e.g., SeeingAI \cite{SeeingAI86:online}, Lookout \cite{LookoutA83:online}) to scan objects.} Our participants discussed various challenges they encountered while using vision apps in the kitchen, such as accurately detecting text on different food packages, controlling when text is (or is not) read out, and navigating visual descriptions. P19 described the positive and negative aspects of using AI-based vision apps to scan food products:

\begin{quote}
    ``...If you buy canned fruit they might actually have a little recipe on the side of the can, and sometimes I try to read that and try to figure that out. And that's kind of an adventure. It's not very accurate at all times, but it's something I like doing just because I can't access such information from other sources...''
\end{quote}

When using apps, participants sometimes struggle to get them \textbf{to only read their desired text, and to not just read any text that shows up in the camera's field of view}. P9 explained:

\begin{figure}[t]
    \centering
    \includegraphics[width=1\columnwidth]{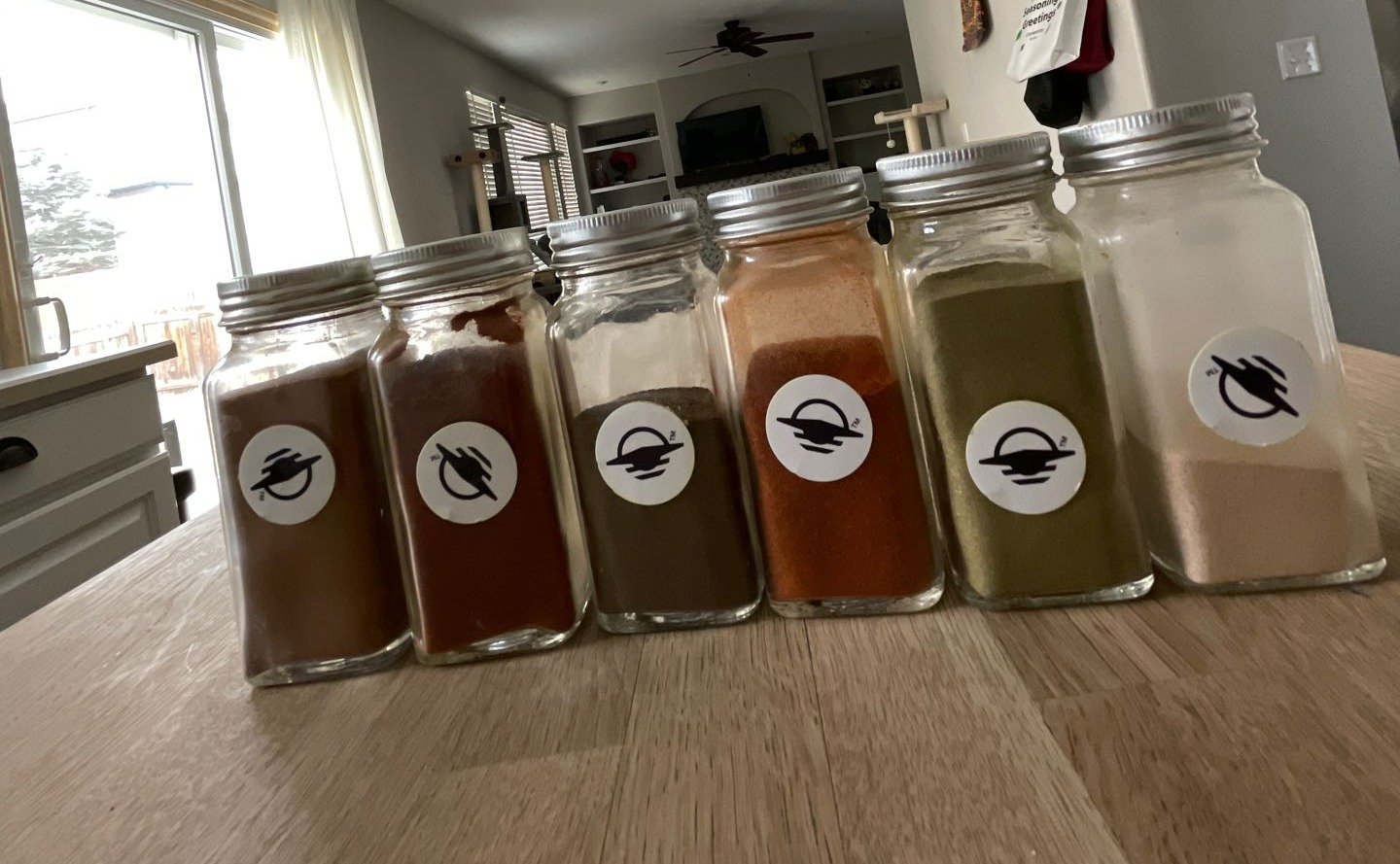}
    \caption{{Participants altered objects to make them more accessible, such as by adding NFC labels to spice jars.}}
    \label{fig:rfid}
    \Description{This figure shows six spice jars with different colors placed in a row on a table. For each jar, there is a white sticker on the front side.}
\end{figure}

\begin{quote}
    ``...So those are not the best if you want to get the ingredients honestly, because it would just start reading whatever is in front of you, but there's no real order or you cannot explore using...So, it would be nice if there was a mode where you can really scan or take a picture of the product and it would give you a better interpretation of columns and better interpretation of tables on a package...''
\end{quote}

P6 desires \textbf{more fine-grained control over what information is presented at what time}, for example describing what each object is when first organizing ingredients, but not to repeat that every time:

\begin{quote}
    ``...I don't want the app to track other things. For example, I only want to know was it like a package of fish, or is it like a package of paddies. I only want to know this information while I'm turning on the system. ''
\end{quote}

Beyond extracting information from physical packages, participants also faced challenges when identifying ingredients in similarly-shaped containers, such as spices. Some participants choose to \textbf{apply tactile markers or NFC tags to distinguish similarly-shaped objects} (e.g., WayAround Tagging System (Figure \ref{fig:rfid}) (P3, P6, P7, P18). 

\subsubsection{Managing Measurements}
Using measuring cups or spoons presented accessibility challenges when the cook must measure something that partially fills the measuring cup (e.g. detecting when a measuring cup is half full). Participants \textbf{favored smaller measuring tools that could be filled completely, so as not to require reading markings on them.} P8 commented:

\begin{quote}
    ``...I frequently employ a variety of measuring tools for different recipes. Depending on whether I'm measuring liquids or solids, I use distinct measuring cups with different units. I tend to favor smaller measuring cups and measure multiple times, as it's more convenient for me than attempting to use a larger measuring cup, which can be challenging to measure a portion of it...''
\end{quote}

Participants found recipes challenging when they mixed units of measurement, as that required using and coordinating more measuring tools. Participants noted that \textbf{automatic unit conversion} could simplify this task (P2, P8, P11, P15).

\begin{table*}[t!]
\resizebox{\columnwidth}{!}{%
\begin{tabular}{l}
\hline
Recipe step (visual descriptions in \textbf{bold})                                                                    \\ \hline
Bake until baklava is \textbf{golden and crisp}.                                                                                                                \\
Drop the covered ice cream balls, into the hot oil and fry, turning occasionally so they \textbf{color evenly, until golden brown on all sides}. \\
Heat 2 tablespoons olive oil over high heat until \textbf{shimmering}.                                                                                          \\
Continue baking until \textbf{lightly browned on top}.                                                                                                          \\
Fold the \textbf{sides, top, and bottom} of each husk in toward the center to enclose dough.                                                                    \\
Add the onions to the skillet and sauté until \textbf{translucent}.                                                                          \\
Slowly knead in a small amount of water at a time, using just enough to form a soft dough that is pliable and \textbf{even in color}.                           \\
Deep fry until \textbf{bright red in color} and crispy.                                                                                                        \\
Reduce heat to 425 degrees F (220 degrees C) and continue baking until pastry is a rich, \textbf{golden brown}.                                                 \\ \hline
\end{tabular}%
}
\caption{Recipe steps gathered for the recipe editing task. Recipe steps often included inaccessible visual descriptions.}
\label{table:Recipesample}
\end{table*}

\subsection{Following Recipes while Cooking}
\label{Reference and Instructional Information Following while Cooking}
\begin{figure}[t]
    \centering
    \includegraphics[width=1\columnwidth]{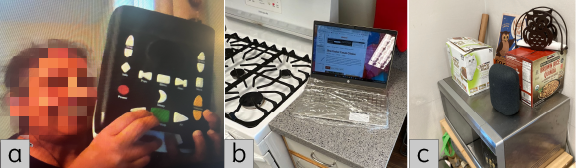}
    \caption{Technology used to accessing recipes. (a) P4 uses an accessible audio player to play back recipes; (b) P2 puts cling wrap on their laptop to prevent it from getting dirty;  (c) P11 uses a smart speaker in the kitchen to access recipes and timers hands-free.}
    \label{fig:device}
    \Description{There are three subfigures. (a) One person is using a tactile-based device by using one hand to hold the device and the other hand to feel the buttons. (b) A laptop with plastic wraps on the keyboard that is placed on the counter close to a gas range. (c) There are several small boxes and a smart speaker placed on top of the microwave oven.}
\end{figure}

Most participants listen to recipes in audio format while cooking. Participants use a variety of tools to access recipes while cooking, including their smartphone-based screen reader (9), laptop screen reader (8), audio players/text-to-speech apps (7), smart speakers (6), physical cookbooks (3), and smartwatches (1) (see figure \ref{fig:device}).

\subsubsection{Navigating Recipes while Cooking}
Participants generally (N=17) used \textbf{touch gestures to navigate while cooking, either via a screen reader or smartphone}. One challenge is that \textbf{screen readers often repeat unnecessary information, even when a cook needs to act quickly}. This problem also occurs when recipes are not broken down into individual steps. P6 explained:

\begin{quote}
    ``...When I'm looking at the steps...I would listen to it. and then go back and complete a couple of actions, and then have to re-listen to the step. [When replaying the step] It often plays from the beginning of the step. And every time, I ended up having to listen to information that isn't necessarily relevant to the actions I'm actually doing...''
\end{quote}

Devices with \textbf{physical buttons, such as dedicated audio players, may be easier to manipulate while cooking} (see Figure \ref{fig:device} (a)). P2 commented: 

\begin{quote}
    ``...I'd have it on my Victor Stream [audio player]. and I could just press the play/pause button and rewind it if I needed to, or fast-forward it...''
\end{quote}

While some participants have \textbf{access to smart speakers, it can be difficult to navigate through a recipe using voice commands alone.} This difficulty is exacerbated by devices that repeat information or forget the current state of the recipe (P3, P15). P15 explained:

\begin{quote}
    ``...I typically do not use Alexa to guide me through the recipe while cooking. It is very common that it does not recognize your wording clearly and loses the current stage of information...''
\end{quote}

\subsubsection{Protecting Devices from Liquids and Other Hazards}

One commonly described challenge was interacting with devices while in the process of cooking, especially when the workspace could be harmful to electronic devices (i.e., wet or messy) or when their hands are dirty. Participants described several strategies for overcoming these challenges:  \textbf{positioning their devices far away from the cooking area}, \textbf{adding protective coverings} such as plastic wrap (see Figure \ref{fig:device} (b)), or \textbf{using voice or other hands-free interaction methods} (e.g., using a smart speaker (see Figure \ref{fig:device} (c))). P2 described his strategy:

\begin{quote}
    ``...When I was using my laptop, I would get like Saran Wrap or something, and I kind of wrap it over the keyboard. So when I was pressing the buttons, I knew it was kind of covered. So I could still press the buttons, but then the Saran Wrap would be covering it...''
\end{quote}

\subsubsection{Memorizing Steps to Reduce Interactions with Recipes}
One strategy that applied regardless of the technology used to follow a recipe was to memorize several steps at a time to \textbf{reduce the number of times they consulted the recipe}. One participant (P7) shared her personal experiences on this matter:

\begin{quote}
    ``...And I look, kind of skim back and forth, and try to keep remembering. `Okay. So this much water for the rice. And no, this time when I make it, I'm not putting ginger in there, but I did last time for the slightly other recipes that they sent me.' So yeah, looking back and forth to figure out what am I throwing in there and when and if my timing is right. So yeah, just scrolling back and forth as I went, as I would do steps...''
\end{quote}




\section{Findings: Recipe Content and Structure (RQ2)}
During the study, participants read along with recipes that were chosen by the research team based on the three dishes they wanted to learn and suggested recipe sources. They provided feedback about their preferred structure, descriptions, and levels of detail for these recipes. Specifically, we engage the question: \textit{What types of information, descriptions, alternatives, and structures might make the cooking process more effective?}
\label{findingRQ2}


\subsection{Recipe Content Preferences}
\label{information addition}

Through reviewing recipes with the research team, participants shared their feedback on what information was necessary, unnecessary, or missing in each step.


\subsubsection{Alternatives for Visual Descriptors}
\label{Non-visual Alternative Descriptions}
Recipes often include steps that require visually inspecting food as it is being prepared, such as checking doneness by examining the food's color (see examples in Table \ref{table:Recipesample}). \textbf{These visual checks are inaccessible to many blind cooks.} One of P13's sample recipes included the instruction: ``Toast split bread under the broiler. Remove bread when it is \textbf{toasted golden brown in color.}'' This accessibility problem was mentioned by most participants as an obstacle that they could not easily overcome. P7 commented:


\begin{quote}
    ``...It is really common to see `cook this until golden brown' in recipes, but it is sometimes hard to standardize the conversion to a description in time or texture. By this time, I always need to further search online for a different recipe for the same cuisine and check if it has a different description...''
\end{quote}

Many recipes include time-based instructions (e.g., ``pan fry for 5 minutes'') along with visual descriptors. However, \textbf{timing information is not always sufficient, as cooking times and equipment vary}. P19 explained: 

\begin{quote}
    ``...Yeah, so just knowing the time of cooking didn't give you how well things cooked. So that's the thing. Sometimes if your water evaporates faster or if your heat is too low, it will be raw on the inside. So, it's not really telling you how you know if they're cooked through. Even after an hour passes, if you have your heat too low, then they might still be raw...''
\end{quote}

Besides timing, our participants highlighted the \textbf{importance of having texture descriptions (4), sound descriptions (6), as well as smell descriptions (3) to substitute the use of visual descriptions}. P4 commented on the importance of having smell descriptions:

\begin{quote}
    ``...Most of the recipes will not have the smell part of the description. They will say, when it looks reddish brown, caramelized onion. So caramelized onion, there will be a smell...Caramelized onion smells sweet and nice...You will know the smell of when it is cooked...Onions will smell differently than cooking potatoes or cooking peas or whatever you are cooking...''
\end{quote}

\subsubsection{Summaries and Outlines}
Participants noted that, before starting cooking, they often read through the entire recipe to ensure that they had the needed ingredients and tools, and to make time to prepare the dish. Participants using screen readers found that \textbf{reading the entire recipe often took too much time, often because the recipe contained extraneous information} such as stories about the recipe. Participants \textbf{prefer having an overview or summary at the beginning of the recipe}. P9 explained her desire to know how the dish would turn out before reading the recipe: 

\begin{quote}
    ``...And I also think getting a short overview of how the recipe is going to turn out. I guess or some of the ingredients, maybe the basic ingredients that you're going to need without telling you the amounts necessarily. But at least that way you can figure out if, oh okay, I do have all the ingredients to make it, or oh, you know what, I'm missing something basic...''
\end{quote}

Furthermore, beyond a high-level overview at the beginning of the recipe, our participants also described their \textbf{desire to have a short summary of the key task and estimated time for each step at the beginning of each instruction step to reduce the effort of navigation}. P12 mentioned:

\begin{quote}
    ``...I would rather [the summary] be part of the steps, and they shouldn't go on too long. Like the scalding, it should be able to describe it pretty shortly. But I would like it to be right there in that step. You don't want to have to go to someplace else to find it, because then you'd have to find your way back to where you were...''
\end{quote}

\subsubsection{Precise Descriptions of Components and Actions}
Participants noted that some recipe descriptions were vague or left out important details, and noted that images within recipes often provide useful contextual information about the size, shape, and color of ingredients, and about how to physically complete cooking steps. Participants noted that \textbf{recipes should describe in text any contextual information that is presented visually, such as the size of an item or how a particular task should be performed.} For example, one of P16's sample recipes stated  ``form into patties and place on the prepared broiler pan or baking sheet;'' P16 commented:

\begin{quote}
    ``...I'd want to know an approximation of the size of the patty...''
\end{quote}

Similarly, some recipes may omit \textbf{details about specific tools to use}, or might expect readers to infer this information from a photo. In response to the instruction ``Heat oil in a large skillet or wok over moderate heat,'' P9 commented

\begin{quote}
    ``...[I don't know] the size of the skillet, because large could mean different things to different people...''
\end{quote}

Recipes with images might also omit \textbf{details about how to perform certain actions, such as distance, intensity, time, and repetition}. In response to the instruction ``place foil around the edges of the crust to protect it from burning,'' P3 asked:

\begin{quote}
    ``...Yep, how far on the edge? Is it over the edge into the pie? Is it the edge that is not covered by the egg mixture? I think I'd want a little more specific on that...''
\end{quote}

\subsubsection{Substitutions or Alternative Instructions}
Recipes often make assumptions about the tools that the user has available. Our participants noted that they may sometimes perform tasks differently than others, such as by using specialized tools (e.g., accessible measuring cups) or cooking in a certain way for accessibility reasons. Twelve participants noted that they prefer recipes that \textbf{provide information about alternatives, such as substituting ingredients or using alternate cooking tools, and optional steps}. P9 commented on the importance of providing substitutions:

\begin{quote}
    ``...I guess it would also be cool to include in the recipe maybe at the end or in the ingredients list, if there's any substitution for any ingredient that you can make if you don't have that ingredient at home or if there's something that you can skip, optional things. I think mostly recipes do tell you if something is optional, but they don't give you any substitutions. For example, if you're trying to make cookies and then you don't have eggs, they don't necessarily give you substitutions for, okay, if you don't have eggs, we recommend you to use this...''
\end{quote}

When given the instruction``Heat oil in a deep fryer to 400 degrees F (200 degrees C). Deep fry samosas until golden brown, 3 to 5 minutes each'', P15 noted that it would be helpful to suggest alternative cooking methods:

\begin{quote}
    ``...I don’t deep fry things because it is hard for me to track the food. What’s an alternative? If there’s a baking alternative, put it right there in the step. like: For deep frying, do this. For baking: do this. For air fryer: do this...''
\end{quote}

\subsubsection{Time Range and Tolerance Details}

Participants preferred instructions that \textbf{provide a time range for each step, and note how these timings might affect the final product}. P6 commented on the uncertainty of timing included in recipes which results in over- or under-cooking food:

\begin{quote}
    ``...there'll be a lot of times where the cooking time won't be, it won't necessarily reflect how long it takes for, say, chicken to cook. It'll be either undercooked, or sometimes they'll go in the opposite direction and in the recipe they'll give instructions for times to cook a chicken that will end up being completely overcooked because they're afraid that somebody's going to undercook it and get sick...''
\end{quote}

P4 commented on the need to have \textit{time ranges} as descriptions: \textit{``Five minutes timing is not enough, and sometimes they can say a range between five to seven minutes, not just five minutes.''} Furthermore, our participants also commented on the value of \textbf{describing time constraints around when certain steps must happen, and which steps can be combined}. P11 commented on their preferences about time management:

\begin{quote}
    ``...I need specifics, I need temperature, I need time, and things that keep time management going. So a recipe where it will give you the instructions to cook pasta and tell you while you're cooking the pasta now you can do something, you can make the sauce. You can do something else within the timeframe, like the 11 minutes it takes to cook pasta...''
\end{quote}






\subsection{Recipe Structure Preferences}
\label{Structure Considerations}

\subsubsection{Separate Processes into Discrete Steps}

All participants mentioned at some point that \textbf{existing recipes often have steps that are too long, or that combine multiple steps into a single instruction.} Participants suggested that \textbf{recipes should be broken into discrete steps for each ingredient and action.} P19 noted that it can be confusing when multiple ingredients are added in a single step:

\begin{quote}
    ``...when you were mixing the eggs white with the milk. that's one step, then, when you fold in the egg yolks. I think that would be another step. That's another that you know. That's another agent entering. So anytime you have to add another ingredient. I would, but I consider that another step...''
\end{quote}

Similarly, P11 noted that each action should be described in a separate step:

\begin{quote}
    ``...You have something, you're putting a bunch of stuff in and cooking it for 25 minutes, that's your first step. And then your next step is to put more stuff in, and cook that for six more minutes. The next step that changes or alters the cooking time or adds to the cooking time, makes that its own step. And then try to keep the activities in the step logically linked. So again, if you're manipulating food, cooking, adjusting heat, timing, that's one step...''
\end{quote}

Recipes that did not follow these rules could be confusing. For example, P2 noted that a recipe was not clear about how often a repeated action should be performed:

\begin{quote}
    ``...No, I do not understand it all. Maybe because it was adding, adding the milk, you know, and then whisking it until it thickened, and then adding more milk and whisking it until thickened. You might wanna break that down. So because you're adding milk. In fact, I can't remember if it was 3 or 4 times. you might want to add each one of those. So if I was doing something new, I'm, you know, might say, okay, I've done that step. Okay, now I'm on step 2, and so the second time I add the milk, and then you know, step 3...''
\end{quote}

\subsubsection{Organizing Ingredients by Use or Type}
P16 noted that \textbf{some recipes list ingredients in confusing order}, and that ingredients should be ordered by when they are used:

\begin{quote}
    ``...Yeah, I would say my ideal recipe is going to be my ingredients list first, grouped together based on usage. So, back to that banana bread, I want the baking soda, the flour, and the salt to be at the top of the list. And then I want it to be sugar, brown sugar, and bananas. And then I want it to be chocolate chips because those are the order that they're added in. And then I like to go into the directions section...''
\end{quote}

Alternatively, some participants preferred that \textbf{ingredients be listed by general category, so that they could be gathered at once and (when possible) mixed ahead of time}. As P3 explained: 
\begin{quote}
    ``...For ingredients, I'd want...A lot of times in baking recipes, wet ingredients together, dry ingredients together...''
\end{quote}

Participants emphasized the importance of \textbf{using consistent language to describe ingredients}, both in the ingredients list and later in the instructions. When navigating a recipe with a screen reader, it is often difficult to jump back and forth between the ingredients list and the current step, as explained by P3:

\begin{quote}
    ``...[is the garlic] dry? Is it dry garlic? Or was it garlic powder? Is it garlic salt? Whatever it is called for in the ingredient list, just call it by the same name all the way through, I'd say. Well, all the spices. Is it brown sugar? Is it white sugar? Is it ... I'm assuming it's ground cinnamon and other things. Just keep consistency between the ingredient list and the step...''
\end{quote}

\subsubsection{Group Steps into Logical Chunks}
Recipes comprise a list of steps in order with the expectation that users will complete all of the steps in order. However, there may be times when \textbf{blind cooks perform steps out of order, or break a recipe into several working sessions due to timing or accessibility constraints}. P6 mentioned his desire for \textbf{recipes that are broken into logical chunks} so that he can work on each part as time permits: 

\begin{quote}
    ``...Making sure to you know if you're putting a whole dish together, and you're preparing a couple of different specific things that are gonna go together. Say, you're cooking a sauce or a gravy, and then separately. You're cooking, you know, meat like a steak, or and then, you know, maybe you're also cooking, you know some onions and peppers to go on top of that clearly separating that into different sections is a big help for recipes...''
\end{quote}

Participants also preferred when \textbf{recipes note which ingredients will be used multiple times}. P7 explained:

\begin{quote}
    ``...Oh, right. So you want two of them in this part and then one of them in that part. If more ingredients were that way, six cloves of garlic divided and then you'd know, Okay, let me just throw these in two separate areas so that I don't accidentally put all of it in one space...''
\end{quote}



\section{Findings: How Technology Can Improve Recipe Accessibility (RQ3)}
\label{findingRQ3}

During our interviews, participants shared their current challenges and frustrations related to following recipes while cooking. These examples led to further discussion about how technology could be improved to mitigate these challenges. Here we consider the question: \textit{How can technology make following recipes more accessible?}

\subsection{Improving Recipe Organization, Navigation, and Level of Detail}

\subsubsection{Provide a Level of Detail that Matches the User}
Participants may prefer different levels of detail depending on their skill or familiarity with the recipe. Since there may be no one-size-fits-all level of detail, \textbf{recipes can be adapted to the user's skills and experience}. P9 explained: 

\begin{quote}
    ``...It has to be detailed, but not so much that you have a paragraph for each step. So I don't want to get all the information, but I do want to get the information that I need. And my background is in culinary arts, so maybe for me, it's easier to understand even with fewer words. But if I learned to cook or started cooking while I was blind, I would definitely want to get some non-visual pointers in my recipes...''
\end{quote}

Participants noted that \textbf{each user might prefer more or less detail for a recipe, depending on their stage within the cooking process}. P1 said:

\begin{quote}
    ``...I'm thinking because I think there are different modes of viewing a recipe. So it could be that when someone is just looking at the recipe or when they're battle planning, they're preparing to cook, that information is annoying, it's too much. It is too much, it's cognitive overload to read all that in one step...''
\end{quote}

\subsubsection{Communicating Information via Voice Characteristics}
Some participants suggested that a system could \textbf{communicate contextual information by changing the characteristics of text-to-speech voices}, such as using different voices to read different parts of the recipe (P12), or allowing users to adjust speed and volume during specific cooking stages (P2, P3, P4, P14, P17, P20). For example, P4 noted that she might use a fast voice when cooking alone, but slow the voice down if she was cooking with a friend.

\subsubsection{Enable Non-Linear Navigation through Recipes}
\label{Status and Time Management}
Participants described \textbf{the difficulty of tracking steps and managing timing, especially when multiple steps [must happen simultaneously]}. P11 explained:

\begin{quote}
    ``...I need time, and things that keep time management going. So a recipe where will give you the instructions to cook pasta and tell you while you're cooking the pasta now you can do something, you can make the sauce. You can do something else within the timeframe, like the 11 minutes it takes to cook pasta...''
\end{quote}

One way to address this challenge is to \textbf{provide alternative paths for stepping through a recipe}. P9 presented an example of her desired navigation strategy:

\begin{quote}
    ``...Yeah. So how it works is I'll ask for a recipe, it will give me the first option. If I like it, I can give it two different commands. The first command is to start with ingredients or start a recipe or something like that. I'm not getting the right wording, but it's the concept of. So when you start gathering the ingredients, it will go one at a time and you just tell like the next ingredient. Of course you have to say her name and then next ingredient, next ingredient. And then it will tell you when you're at the end of the ingredients, and it will ask you if you want to continue with the instructions and then it goes one step at a time...''
\end{quote}

\subsection{Supporting Users in Learning to Cook}
We asked the adaptive cooking instructors about how their students work when taking classes, and how they could be better supported in their cooking activities. 
\subsubsection{Customizing Instructions for Skills and Ability}
\label{Customized Instruction Based on Skills and Ability}
All instructors mentioned the importance of adapting instructions to the student, depending on their age, cooking experience, skill, and visual abilities. For example, I3 noted that some students knew how to cook before the vision loss and took cooking classes only to re-learn cooking skills non-visually, while others are learning to cook for the first time. We also found that people with fewer years of cooking experience or who have never cooked before blindness (e.g., P16) are more likely to prefer more detailed cooking descriptions and links to external resources through technology than people who had many years of cooking experience before blindness (e.g., P6). I4 emphasized the importance of \textbf{providing customized cooking instructions based on individual background}:

\begin{quote}
    ``...I found these newly blind people who show interest in cooking tend to have different understandings of cooking. Emmm, I remember there was a student who was a chef before becoming blind. His knowledge was much stronger than mine, so I only needed to teach him the basics of using tools non-visually, etc. But for many many other students who were blind and had never cooked independently before, I need to start from scratch to teach them everything...''
\end{quote}

As noted in Section \ref{Instructional information extraction and manipulation}, more experienced cooks often edited recipes while transcribing them into a text file. Essentially, these cooks are customizing their own recipes, although \textbf{customizing and editing recipes could be made easier}. P8 described how and when he chooses to customize recipes:

\begin{quote}
    ``...Yeah, my first time when following the recipe, I just get to follow the recipe. And wherever I feel, okay, I'm not comfortable with this, I can actually make my own personal adjustments. Yeah. Yeah. That's just my own recipe...''
\end{quote}

The process of customizing a recipe also \textbf{creates an accessible version of the recipe that may be useful to other blind cooks.} P7 commented: \textit{``I wish I could just see and use other people's edits on the recipe who are also blind, which could save a lot of time. Other people could then see my edits too.''}

\subsubsection{Structured Discovery and Professional Instructions}
Instructors mentioned the teaching approach called Structured Discovery as a way to support students with varying skills and abilities. In Structured Discovery, instructors refrain from using hand-over-hand guidance and encourage students to experience cooking skills on their own with verbal guidance from instructors. I1 explained:

\begin{quote}
    ``...We're a structured discovery center, so most of the time it's students having to work through these things on their own...Typically, I don't do a lot of hand-over-hand teaching. If I'm giving instruction or if students need any clarification or they're learning something new, I always try to give verbal instructions first...''
\end{quote}

We also found that our participants prefer \textbf{instructions that are intended for blind cooks specifically, rather than for the general population.} P11 commented on the usefulness of BeMyEyes \cite{williams2013preliminary} in his cooking:

\begin{quote}
    ``...So if they had some kind of connection to chefs or anyone that can note that they are cooks or they have good cooking knowledge, then when you go in and you specifically request. Hey, I need some help in the kitchen, then you can get someone who's knowledgeable about cooking methods or just things that you need a little bit of guidance or things that you need to do something in the kitchen...''
\end{quote}

\subsubsection{Answering Questions and Providing Assurance}
\label{Intended Verbal QA}

Our instructors (I1 - I4) mentioned that during cooking classes, their students \textbf{constantly ask questions for confirmation and learning purposes}. They noted that \textbf{asking questions while practicing can be a distraction, and that students must learn to act without external confirmation.} I3 explained:

\begin{quote}
    ``...A lot of times it's just looking for reassurance. Did I do this right? That's the biggest one. I'm not sure if I'm doing this right or if this is the right piece of equipment. Because we try to get that out of the way ahead of time. I try to make them think through the process and ask most of the questions that they would've asked in the process ahead of time in conversation. It takes a lot of time to do that, but it saves time during the actual cooking process...''
\end{quote}

Our participants mentioned that existing ways of asking smart speakers, searching through recipes, or using vision assistance Apps either cannot get the expected answer or the whole process takes too long (Section \ref{Reference and Instructional Information Following while Cooking}). We found that cooking-related questions are often very context-specific, such as ``how long should I cook this tomato again?'' This points to the \textbf{importance of understanding the question's intent and the current stage of the recipe.} P3 shared some common questions that she has while cooking:

\begin{quote}
    ``...Yeah. Standard questions like, what's the cooking time? What's the oven temperature? Things like that. That would be handy to know right away...''
\end{quote}

\subsection{Integrating Technology into Kitchen Activities}
As discussed previously, participants often struggled with using their normal cooking devices in the kitchen, and invented workarounds for doing so, such as covering their laptop with cling film. During our interviews, we discussed how technology could better integrate into kitchen activities.

\subsubsection{Using Multiple Devices Together}
\label{Form Factors and Multimodal Access}
One problem noted by participants was the \textbf{difficulty in sharing information across multiple devices}, such as when they saved recipes on a computer and read them on a mobile device. P12 suggested that cooking information should propagate across his computing devices, including his smartphone and smartwatch:

\begin{quote}
    ``...[it should] notify me on my phone and my Apple Watch when it is done by vibration or sound whenever it is appropriate...''
\end{quote}

\subsubsection{Integrating At-School Learning and At-Home Practice}

I1 noted that \textbf{recipes and tools used at school should also be available at home}, to support practicing skills:

\begin{quote}
    ``...Yeah, a huge, huge thing that I would love to support is having as many smart devices in the house. However, if a person is in training and they have the smart devices in the kitchen and they don't have them at their house, then that's going to be a challenge. So I also think meeting the students where they're at is a good thing, too. So meaning that besides the smart devices, if they don't have them at home, we'll then also teach them without those smart devices...''
\end{quote}

Instructors mentioned that their students may \textbf{struggle to reproduce work from class at home}, with different tools available and without feedback from instructors (I1, I2, I4). I4 explained:

\begin{quote}
    ``...I mean, many of my students do not have the same set of tools or devices at home compared with the kitchen at the rehab center, they sometimes struggle with the tools they have at home and there is not much I could do to help them...I usually have my phone available after hours and try to clarify and help as much as I can, but it is usually hard to do so through phone calls...''
\end{quote}

\subsubsection{Supporting Hands-Free Interaction}
In a previous section, we noted the challenge of switching between cooking activities and interacting with technology. Participants suggested that technology could enable \textbf{hands-free interactions with their recipe devices}, would reduce the need to rinse and dry their hands during cooking. P19 commented:

\begin{quote}
    ``...It would be nice to have either something that can be on some sort of a stand or hands-free something that you can ask your device to do, especially when you have your hands full. So you got your potatoes in one thing, and you got your stuff, and it's really difficult to actually grab your phone and try to see. You know. Rinse your hands and then get back to other stuff. It'd be really nice to have something hands-free that can do that or even a wearable that concentrates on stuff like that cooking...''
\end{quote}

Alternatively, computing devices could \textbf{provide simplified touch interfaces to prevent mistaken input.} P1 said:

\begin{quote}
    ``...So sometimes with apps, you can restrict the gestures that are available to make so that, what am I trying to say? When you're navigating with a screen reader on the phone, you have all the gestures available to you so you can accidentally go out of the file or you can accidentally lose your place. So if it was an app...that would restrict the gestures and only leave you a limited number of gestures available in that app. So you can't accidentally go out of the recipe...''
\end{quote}

Our participants also emphasized the importance of \textbf{leveraging multiple interaction modalities while cooking} (P2, P8, P11, P17). P17 commented:

\begin{quote}
    ``...I only have one option to interact with my recipe reader which is through touch, I had to wash my hands multiple times and I wish I could simply use voice or wave my hand to move the recipe to the next step...''
\end{quote}


\section{Discussion and Future Directions}
\label{Discussion}

In this section, we will discuss opportunities and considerations to augment recipe information and interactive technologies to better support non-visual access to cooking instructions and related information. 


\subsection{Augmenting Recipes to Make Them More Accessible}

The first strategy we suggest is to augment the information content within the recipe itself. This would include modifying recipes to add the suggested information additions or to make them more widely adopted. Referring back to the previous section, this would include support for non-visual alternative descriptions for objects (including sound, smell, or touch) and especially changes in ingredient or cooking status (Section \ref{information addition}). Incorporating both summaries and additional precision where appropriate. Today's AI tools (e.g., Large Language Models (LLMs)) may be used to summarize short pieces of the recipe to achieve the vision of concise step descriptions based on key actions without the need to re-write the full recipe \cite{pini2019ai,van2021using,liu2024selenite}. By gathering further information from blind cooks, it may also be possible to identify unique substitutions for actions, in addition to utilizing resources for ingredient substitutions. Furthermore, leveraging the expert knowledge of chefs and food safety experts we may also be able to identify resources to build a repository of tolerance information for various ingredients.

\subsubsection{Non-visual Database for Common Cooking References and Substitutions}
In our study, we showed that our participants prefer having non-visual interpretations to substitute visual descriptors (e.g., cook until golden brown) to reduce the barriers to learning and following cooking instructions effectively. We found that our participants gradually supplemented their cognition by either converting the visual description of a certain action or process non-visually before starting to cook, or utilizing strategies directly learned from blind cooking classes. This approach varies based on different ingredients, ways of cooking, size of the food, shape, and many other variables, which makes the process very experimental. We believe that it could be beneficial to create a database of non-visual descriptors for common cooking tasks and references to allow simple and automatic substitutions of relevant information. This database of recipe descriptions could include attributes such as [Original Visual Description], [Ingredient], [Cuisine Name], [Cooking Method], and [Non-Visual Description]. We believe this would reduce the learning curve for people with vision impairments, and having such a database could also contribute to recipe customization models (e.g., \cite{teng2011reciperecommendation,chang2018recipescape,kusu2017calculating}) to automatically transform recipes with non-visual equivalence.

\subsection{Facilitating Information Extraction for Kitchen Tools and Ingredients}

\subsubsection{Automatic Extraction of Visual Content from Photos and Video}
From our study, we found existing online recipes contain various visual content that are inaccessible to blind people, such as having pictures or videos attached to the text to show the ingredients, steps, or size of objects. For example, in a video recipe, the person might verbally say, ``Cook this until golden brown,'' however, we can leverage data extraction tools \cite{wong2003new} to identify any text shown in the video that indicates the timing or even extract relevant or unique sounds that is generated through the process (e.g., sizzling). This approach could potentially contribute to the non-visual reference database for visually impaired people to leverage while following a typical recipe.

\subsubsection{Supporting Extraction of Static Information from Physical Objects}
In our paper, we found that our participants leveraged vision apps to identify information on physical packages, and we showed various challenges that they encountered through the process (Section \ref{information identification}). This is similar to prior research that explored package fetching by people with vision impairments \cite{lei2022shake}. Here, we propose future research to consider providing universal designs of packaging for visually impaired people, which includes considerations of the absolute position of certain information on the physical package (e.g., nutrition list, ingredients list, cooking instructions), support recognition of unpackaged items (e.g., a bunch of parsley from the store), as well as tactile readings \cite{ribeiro2019information}. Due to the existing difficulties of using vision Apps to identify visualizations (Section \ref{information identification}), we also recommend package designers to utilize plain text structure and provide high-level summaries for easier reading through vision apps.

\subsubsection{Supporting Extraction of Dynamic Information from Physical Objects}
Furthermore, objects used in the kitchen often change or move, thus necessitating a constant re-evaluation of space and the objects that are being used. Future technology tools need to be able to support identifying these kinds of changes and communicate such information at appropriate stages during the cooking process \cite{li2019fmt}. Our participants mentioned how the granularity of recipe information may change through different cooking stages (Section \ref{findingRQ2}), and the same applies to details about physical objects, especially while learning (Section \ref{findingRQ3}). For instance, consider a simple example of a knife, a versatile tool in the kitchen. At the start of a cooking task, people might need information about its location to ensure it's safely stowed in its designated spot. However, as the cooking progresses, the context for this object evolves. People may need updates on whether the knife has been used, requiring information on its cleanliness and any food residues. Later in the process, people might require data on the knife's orientation, particularly if you've set it aside with the blade exposed. This need for dynamic information motivates technology to assist blind cooks by reducing cognitive load and increasing environmental awareness.


\subsection{Supporting More Accessible Kitchen Work}

\subsubsection{Multimodal Opportunities for Blind People in the Kitchen}
We also learned that our participants prefer having the option to have multiple devices (e.g., connect the smartphone with the smart speaker) and modalities (e.g., sound, haptic) in the kitchen to interact with recipes while cooking for convenience and reduce false-positives (Section \ref{Form Factors and Multimodal Access}). This would provide the freedom to choose the modality to interact with the system and receive output in preferred ways \cite{li2022freedom}. This provides opportunities to consider: 1) How the input and output modalities should be designed? 2) What are the preferred gestures for the multimodal systems? 3) what are the design considerations of having multimodal systems in the kitchen for blind people (e.g., deployment, activation)? 4) What are the broader perceived benefits beyond interacting with recipes by using multimodal systems in the kitchen?

\subsubsection{Information Modality for Cognition}
In our paper, we showed that our participants leveraged text and video for different purposes (e.g., manually transcribing video content for details to text). Specifically, we discussed how our participants prefer leveraging sound and sayings from the cook to gain motivation for cooking (Section \ref{information seeking}). Beyond cooking, our findings correlate to prior research on leveraging multimodal guides or interaction to support blind people in everyday activities, such as navigation \cite{lin2014context}, social interaction \cite{kianpisheh2019face}, game \cite{sanchez2015multimodal}, doing makeup \cite{li2022feels} and art experiences \cite{cavazos2018interactive}. Therefore, to enhance the adaptation of information, further research should also consider how to enable information in different modalities (e.g., sound, smell, tactile) to better motivate and assist people with vision impairments in accessing information and interacting with devices in cooking.

\subsubsection{Conversational User Interfaces for Kitchen-AI Systems}
In our findings, we showed that our participants commonly leveraged communications with their instructors in the cooking class to reassure their process and ask for clarifications on specific steps (Section \ref{Intended Verbal QA}). This provides opportunities for leveraging tools such as LLMs \cite{kasneci2023chatgpt} to provide prompts to answer questions that are highly contextual based on specific recipes to support learning experiences and agency to cook without instructors. For example, this includes providing details of specific steps that require high memory load (e.g., quantity, time, ingredients), having detailed guidance on how to perform certain actions without vision (e.g., suggestions on measuring food doneness), having summarization of a recipe, and describing alternative ways to do something. To do so, future research should conduct contextual inquiries to observe existing Q\&A conversations between blind cooking students and their instructors in cooking classes to further explore the proper prompts to provide accurate and ethical responses \cite{kasneci2023chatgpt,zhou2023ethical} through the conversational user interfaces. 

\subsubsection{Customization Based on Individual Backgrounds and Conditions}
In our findings, we showed that our instructors tailored their instructions based on people's diverse backgrounds (Section \ref{Customized Instruction Based on Skills and Ability}), such as people with or without cooking experiences before becoming blind or people of different ages \cite{kuoppamaki2021designing}. In our study with blind cooks, we had participants who were totally blind as well as legally blind who were capable of finger counting (Table \ref{table:participants}). By analyzing their behaviors, we found that our participants who are legally blind and can perform finger counting or exhibited some degree of tunnel vision (e.g., P6, P9, P12) also employed vision-assistive technologies during cooking procedures non-visually, such as P6 used BeMyEyes to read physical packages, P9 used SeeingAI to read labels, and P12 used voiceover to walk through recipes on mobile devices, which is identical to other participants who are totally blind. On the other hand, we found there are slight differences regarding how they incorporated vision-assistive technologies with their remaining visual capabilities (e.g., light perception). For instance, while using SeeingAI to identify object locations, P12 utilized his light perceptions to estimate the facing direction based on the daylight from the window and the surrounding area that might contain certain objects (e.g., milk in the fridge). Drawing insights from both blind cooks and instructors, our recommendation for future systems is to offer users comprehensive control options. This includes preferences for spoken information, customizable levels of navigational assistance, and targeted guidance on specific cooking skills. Empowering end-users with detailed controls can potentially enhance their individual user experience and adaptability to diverse needs.

\section{Limitations}
In our semi-structured interview, we only recruited four cooking instructors from a vision rehabilitation center. We found different people have different ways of instructing visually impaired people in cooking activities. In future research, we believe it would be useful to conduct interviews with different people who have different relationships with blind cooks (e.g., parents, friends, cooking instructors) to understand the practices and challenges that they have to provide proper support to people with vision impairments. Furthermore, we only conducted studies with people who are legally or totally blind and who have cooking experience. People with low vision that have different vision conditions (e.g., blurry vision, central vision loss) or people who just started learning how to cook, might have different practices and challenges of accessing recipe information while cooking. Future research could conduct comparison studies to understand the differences in individual needs and preferences regarding instructional information by people with different vision conditions, cooking experiences, and different cultural backgrounds \cite{li2021choose,vashistha2015social}. Moreover, all of our participants used certain forms of technology to access and interact with recipe information. Future work should formally evaluate how familiarity with using technology can impact the cooking experience with recipes.

\section{Conclusion}
In this paper, we first presented the methods, strategies, and challenges of accessing recipe information at three different cooking stages for people with vision impairments (e.g., making customized edits to ingredients and steps for accessible experiences). We then showed the desired recipe information descriptions and structures that our participants prefer to have while cooking (e.g., non-visual alternative descriptions, improving browsable structure). We also described design features for technologies to better support recipe access (e.g., adaptive reading and voice modulation). We concluded with a discussion of future directions and opportunities for augmenting recipe information, facilitating information extraction for kitchen tools and ingredients, and supporting accessible interactions in the kitchen. These opportunities include novel directions for creating a non-visual database for common cooking references and substitutions, as well as multimodal opportunities for people with vision impairments in the kitchen to support agency and reduce physical and mental effort. Overall, our research provides nuanced insights and design guidelines to enhance recipe accessibility for people with vision impairments.

\begin{acks} 
This work was funded in part by Google Research and NSERC Fellowship PGS-D-557565-2021. We especially thank NFB and SAVVI for supporting participant recruitment.
\end{acks}

\bibliographystyle{ACM-Reference-Format}
\bibliography{main}


\begin{thebibliography}{81}


\ifx \showCODEN    \undefined \def \showCODEN     #1{\unskip}     \fi
\ifx \showDOI      \undefined \def \showDOI       #1{#1}\fi
\ifx \showISBNx    \undefined \def \showISBNx     #1{\unskip}     \fi
\ifx \showISBNxiii \undefined \def \showISBNxiii  #1{\unskip}     \fi
\ifx \showISSN     \undefined \def \showISSN      #1{\unskip}     \fi
\ifx \showLCCN     \undefined \def \showLCCN      #1{\unskip}     \fi
\ifx \shownote     \undefined \def \shownote      #1{#1}          \fi
\ifx \showarticletitle \undefined \def \showarticletitle #1{#1}   \fi
\ifx \showURL      \undefined \def \showURL       {\relax}        \fi
\providecommand\bibfield[2]{#2}
\providecommand\bibinfo[2]{#2}
\providecommand\natexlab[1]{#1}
\providecommand\showeprint[2][]{arXiv:#2}

\bibitem[\protect\citeauthoryear{Abdolrahmani, Kuber, and Branham}{Abdolrahmani et~al\mbox{.}}{2018}]%
        {abdolrahmani2018siri}
\bibfield{author}{\bibinfo{person}{Ali Abdolrahmani}, \bibinfo{person}{Ravi Kuber}, {and} \bibinfo{person}{Stacy~M Branham}.} \bibinfo{year}{2018}\natexlab{}.
\newblock \showarticletitle{" Siri Talks at You" An Empirical Investigation of Voice-Activated Personal Assistant (VAPA) Usage by Individuals Who Are Blind}. In \bibinfo{booktitle}{\emph{Proceedings of the 20th International ACM SIGACCESS Conference on Computers and Accessibility}}. \bibinfo{pages}{249--258}.
\newblock


\bibitem[\protect\citeauthoryear{Adetoro}{Adetoro}{2016}]%
        {niran2016access}
\bibfield{author}{\bibinfo{person}{Niran Adetoro}.} \bibinfo{year}{2016}\natexlab{}.
\newblock \bibinfo{booktitle}{\emph{INFORMATION ACCESS FOR THE VISUALLY IMPAIRED IN THE DIGITAL AGE}}.
\newblock


\bibitem[\protect\citeauthoryear{Alexa}{Alexa}{2023}]%
        {AmazonAl14:online}
\bibfield{author}{\bibinfo{person}{Alexa}.} \bibinfo{year}{2023}\natexlab{}.
\newblock \bibinfo{title}{Amazon Alexa}.
\newblock \bibinfo{howpublished}{\url{https://alexa.amazon.com/}}.
\newblock
\newblock
\shownote{(Accessed on 05/11/2023).}


\bibitem[\protect\citeauthoryear{Allrecipes}{Allrecipes}{2023}]%
        {Allrecip74:online}
\bibfield{author}{\bibinfo{person}{Allrecipes}.} \bibinfo{year}{2023}\natexlab{}.
\newblock \bibinfo{title}{Allrecipes | Recipes, How-Tos, Videos and More}.
\newblock \bibinfo{howpublished}{\url{https://www.allrecipes.com/}}.
\newblock
\newblock
\shownote{(Accessed on 04/18/2023).}


\bibitem[\protect\citeauthoryear{Azenkot and Lee}{Azenkot and Lee}{2013}]%
        {azenkot2013exploring}
\bibfield{author}{\bibinfo{person}{Shiri Azenkot} {and} \bibinfo{person}{Nicole~B Lee}.} \bibinfo{year}{2013}\natexlab{}.
\newblock \showarticletitle{Exploring the use of speech input by blind people on mobile devices}. In \bibinfo{booktitle}{\emph{Proceedings of the 15th international ACM SIGACCESS conference on computers and accessibility}}. \bibinfo{pages}{1--8}.
\newblock


\bibitem[\protect\citeauthoryear{Azenkot, Rector, Ladner, and Wobbrock}{Azenkot et~al\mbox{.}}{2012}]%
        {azenkot2012passchords}
\bibfield{author}{\bibinfo{person}{Shiri Azenkot}, \bibinfo{person}{Kyle Rector}, \bibinfo{person}{Richard Ladner}, {and} \bibinfo{person}{Jacob Wobbrock}.} \bibinfo{year}{2012}\natexlab{}.
\newblock \showarticletitle{PassChords: secure multi-touch authentication for blind people}. In \bibinfo{booktitle}{\emph{Proceedings of the 14th international ACM SIGACCESS conference on Computers and accessibility}}. \bibinfo{pages}{159--166}.
\newblock


\bibitem[\protect\citeauthoryear{Bigham, Jayant, Ji, Little, Miller, Miller, Miller, Tatarowicz, White, White, et~al\mbox{.}}{Bigham et~al\mbox{.}}{2010}]%
        {bigham2010vizwiz}
\bibfield{author}{\bibinfo{person}{Jeffrey~P Bigham}, \bibinfo{person}{Chandrika Jayant}, \bibinfo{person}{Hanjie Ji}, \bibinfo{person}{Greg Little}, \bibinfo{person}{Andrew Miller}, \bibinfo{person}{Robert~C Miller}, \bibinfo{person}{Robin Miller}, \bibinfo{person}{Aubrey Tatarowicz}, \bibinfo{person}{Brandyn White}, \bibinfo{person}{Samual White}, {et~al\mbox{.}}} \bibinfo{year}{2010}\natexlab{}.
\newblock \showarticletitle{Vizwiz: nearly real-time answers to visual questions}. In \bibinfo{booktitle}{\emph{Proceedings of the 23nd annual ACM symposium on User interface software and technology}}. \bibinfo{pages}{333--342}.
\newblock


\bibitem[\protect\citeauthoryear{Bilyk, Sontrop, Chapman, Barr, and Mamer}{Bilyk et~al\mbox{.}}{2009}]%
        {bilyk2009food}
\bibfield{author}{\bibinfo{person}{Marie~Claire Bilyk}, \bibinfo{person}{Jessica~M Sontrop}, \bibinfo{person}{Gwen~E Chapman}, \bibinfo{person}{Susan~I Barr}, {and} \bibinfo{person}{Linda Mamer}.} \bibinfo{year}{2009}\natexlab{}.
\newblock \showarticletitle{Food experiences and eating patterns of visually impaired and blind people}.
\newblock \bibinfo{journal}{\emph{Canadian Journal of Dietetic practice and research}} \bibinfo{volume}{70}, \bibinfo{number}{1} (\bibinfo{year}{2009}), \bibinfo{pages}{13--18}.
\newblock


\bibitem[\protect\citeauthoryear{Blasco, Marco, Casas, Cirujano, and Picking}{Blasco et~al\mbox{.}}{2014}]%
        {blasco2014smart}
\bibfield{author}{\bibinfo{person}{Rub{\'e}n Blasco}, \bibinfo{person}{{\'A}lvaro Marco}, \bibinfo{person}{Roberto Casas}, \bibinfo{person}{Diego Cirujano}, {and} \bibinfo{person}{Richard Picking}.} \bibinfo{year}{2014}\natexlab{}.
\newblock \showarticletitle{A smart kitchen for ambient assisted living}.
\newblock \bibinfo{journal}{\emph{Sensors}} \bibinfo{volume}{14}, \bibinfo{number}{1} (\bibinfo{year}{2014}), \bibinfo{pages}{1629--1653}.
\newblock


\bibitem[\protect\citeauthoryear{Blenkhorn}{Blenkhorn}{1995}]%
        {blenkhorn1995producing}
\bibfield{author}{\bibinfo{person}{Paul Blenkhorn}.} \bibinfo{year}{1995}\natexlab{}.
\newblock \showarticletitle{Producing a text-to-speech synthesizer for use by blind people}.
\newblock \bibinfo{journal}{\emph{Extra-ordinary Human-Computer Interaction: Interfaces for Users with Disabilities, ADN Edwards (ed.), Cambridge University Press, New York}} (\bibinfo{year}{1995}), \bibinfo{pages}{307--314}.
\newblock


\bibitem[\protect\citeauthoryear{Borghini}{Borghini}{2015}]%
        {borghini2015recipe}
\bibfield{author}{\bibinfo{person}{Andrea Borghini}.} \bibinfo{year}{2015}\natexlab{}.
\newblock \showarticletitle{What is a Recipe?}
\newblock \bibinfo{journal}{\emph{Journal of Agricultural and Environmental Ethics}}  \bibinfo{volume}{28} (\bibinfo{year}{2015}), \bibinfo{pages}{719--738}.
\newblock


\bibitem[\protect\citeauthoryear{Branham and Mukkath~Roy}{Branham and Mukkath~Roy}{2019}]%
        {branham2019reading}
\bibfield{author}{\bibinfo{person}{Stacy~M Branham} {and} \bibinfo{person}{Antony~Rishin Mukkath~Roy}.} \bibinfo{year}{2019}\natexlab{}.
\newblock \showarticletitle{Reading between the guidelines: How commercial voice assistant guidelines hinder accessibility for blind users}. In \bibinfo{booktitle}{\emph{The 21st International ACM SIGACCESS Conference on Computers and Accessibility}}. \bibinfo{pages}{446--458}.
\newblock


\bibitem[\protect\citeauthoryear{Cavazos~Quero, Iranzo~Bartolom{\'e}, Lee, Han, Kim, and Cho}{Cavazos~Quero et~al\mbox{.}}{2018}]%
        {cavazos2018interactive}
\bibfield{author}{\bibinfo{person}{Luis Cavazos~Quero}, \bibinfo{person}{Jorge Iranzo~Bartolom{\'e}}, \bibinfo{person}{Seonggu Lee}, \bibinfo{person}{En Han}, \bibinfo{person}{Sunhee Kim}, {and} \bibinfo{person}{Jundong Cho}.} \bibinfo{year}{2018}\natexlab{}.
\newblock \showarticletitle{An interactive multimodal guide to improve art accessibility for blind people}. In \bibinfo{booktitle}{\emph{Proceedings of the 20th International ACM SIGACCESS Conference on Computers and Accessibility}}. \bibinfo{pages}{346--348}.
\newblock


\bibitem[\protect\citeauthoryear{Chang, Guillain, Jung, Hare, Kim, and Agrawala}{Chang et~al\mbox{.}}{2018}]%
        {chang2018recipescape}
\bibfield{author}{\bibinfo{person}{Minsuk Chang}, \bibinfo{person}{L{\'e}onore~V Guillain}, \bibinfo{person}{Hyeungshik Jung}, \bibinfo{person}{Vivian~M Hare}, \bibinfo{person}{Juho Kim}, {and} \bibinfo{person}{Maneesh Agrawala}.} \bibinfo{year}{2018}\natexlab{}.
\newblock \showarticletitle{Recipescape: An interactive tool for analyzing cooking instructions at scale}. In \bibinfo{booktitle}{\emph{Proceedings of the 2018 CHI conference on human factors in computing systems}}. \bibinfo{pages}{1--12}.
\newblock


\bibitem[\protect\citeauthoryear{Charmaz}{Charmaz}{2006}]%
        {charmaz2006constructing}
\bibfield{author}{\bibinfo{person}{Kathy Charmaz}.} \bibinfo{year}{2006}\natexlab{}.
\newblock \bibinfo{booktitle}{\emph{Constructing grounded theory: A practical guide through qualitative analysis}}.
\newblock \bibinfo{publisher}{sage}.
\newblock


\bibitem[\protect\citeauthoryear{Cooper}{Cooper}{2015}]%
        {cooper2015cooking}
\bibfield{author}{\bibinfo{person}{Jenny Cooper}.} \bibinfo{year}{2015}\natexlab{}.
\newblock \showarticletitle{Cooking trends among millennials: Welcome to the digital kitchen}.
\newblock \bibinfo{journal}{\emph{Think with Google}} (\bibinfo{year}{2015}), \bibinfo{pages}{18}.
\newblock


\bibitem[\protect\citeauthoryear{Corp}{Corp}{2021}]%
        {HomeAira36:online}
\bibfield{author}{\bibinfo{person}{Aira~Tech Corp}.} \bibinfo{year}{2021}\natexlab{}.
\newblock \bibinfo{title}{Home - Aira : Aira}.
\newblock \bibinfo{howpublished}{\url{https://aira.io/}}.
\newblock
\newblock
\shownote{(Accessed on 04/08/2021).}


\bibitem[\protect\citeauthoryear{Davis, Nansen, Vetere, Robertson, Brereton, Durick, and Vaisutis}{Davis et~al\mbox{.}}{2014}]%
        {davis2014homemade}
\bibfield{author}{\bibinfo{person}{Hilary Davis}, \bibinfo{person}{Bjorn Nansen}, \bibinfo{person}{Frank Vetere}, \bibinfo{person}{Toni Robertson}, \bibinfo{person}{Margot Brereton}, \bibinfo{person}{Jeannette Durick}, {and} \bibinfo{person}{Kate Vaisutis}.} \bibinfo{year}{2014}\natexlab{}.
\newblock \showarticletitle{Homemade cookbooks: a recipe for sharing}. In \bibinfo{booktitle}{\emph{Proceedings of the 2014 conference on Designing interactive systems}}. \bibinfo{pages}{73--82}.
\newblock


\bibitem[\protect\citeauthoryear{Feiz, Billah, Ashok, Shilkrot, and Ramakrishnan}{Feiz et~al\mbox{.}}{2019}]%
        {feiz2019towards}
\bibfield{author}{\bibinfo{person}{Shirin Feiz}, \bibinfo{person}{Syed~Masum Billah}, \bibinfo{person}{Vikas Ashok}, \bibinfo{person}{Roy Shilkrot}, {and} \bibinfo{person}{IV Ramakrishnan}.} \bibinfo{year}{2019}\natexlab{}.
\newblock \showarticletitle{Towards enabling blind people to independently write on printed forms}. In \bibinfo{booktitle}{\emph{Proceedings of the 2019 CHI Conference on Human Factors in Computing Systems}}. \bibinfo{pages}{1--12}.
\newblock


\bibitem[\protect\citeauthoryear{Foster}{Foster}{2023}]%
        {BowTiePa2:online}
\bibfield{author}{\bibinfo{person}{Kelli Foster}.} \bibinfo{year}{2023}\natexlab{}.
\newblock \bibinfo{title}{Bow Tie Pasta Salad Recipe (with Tomatoes) | Kitchn}.
\newblock \bibinfo{howpublished}{\url{https://www.thekitchn.com/bow-tie-pasta-salad-recipe-23533554}}.
\newblock
\newblock
\shownote{(Accessed on 05/11/2023).}


\bibitem[\protect\citeauthoryear{Fusco, Tekin, Ladner, and Coughlan}{Fusco et~al\mbox{.}}{2014}]%
        {fusco2014using}
\bibfield{author}{\bibinfo{person}{Giovanni Fusco}, \bibinfo{person}{Ender Tekin}, \bibinfo{person}{Richard~E Ladner}, {and} \bibinfo{person}{James~M Coughlan}.} \bibinfo{year}{2014}\natexlab{}.
\newblock \showarticletitle{Using computer vision to access appliance displays}. In \bibinfo{booktitle}{\emph{Proceedings of the 16th international ACM SIGACCESS conference on Computers \& accessibility}}. \bibinfo{pages}{281--282}.
\newblock


\bibitem[\protect\citeauthoryear{Gonzalez, Collins, Azenkot, and Bennett}{Gonzalez et~al\mbox{.}}{2024}]%
        {gonzalez2024investigating}
\bibfield{author}{\bibinfo{person}{Ricardo Gonzalez}, \bibinfo{person}{Jazmin Collins}, \bibinfo{person}{Shiri Azenkot}, {and} \bibinfo{person}{Cynthia Bennett}.} \bibinfo{year}{2024}\natexlab{}.
\newblock \bibinfo{title}{Investigating Use Cases of AI-Powered Scene Description Applications for Blind and Low Vision People}.
\newblock
\newblock
\showeprint[arxiv]{2403.15604}~[cs.HC]


\bibitem[\protect\citeauthoryear{Google}{Google}{[n.d.]a}]%
        {GoogleFo84:online}
\bibfield{author}{\bibinfo{person}{Google}.} \bibinfo{year}{[n.d.]}\natexlab{a}.
\newblock \bibinfo{title}{Google Forms: Online Form Creator | Google Workspace}.
\newblock \bibinfo{howpublished}{\url{https://www.google.com/forms/about/}}.
\newblock
\newblock
\shownote{(Accessed on 08/24/2022).}


\bibitem[\protect\citeauthoryear{Google}{Google}{[n.d.]b}]%
        {LookoutA83:online}
\bibfield{author}{\bibinfo{person}{Google}.} \bibinfo{year}{[n.d.]}\natexlab{b}.
\newblock \bibinfo{title}{Lookout - Assisted vision - Apps on Google Play}.
\newblock \bibinfo{howpublished}{\url{https://play.google.com/store/apps/details? id=com.google.android.apps.accessibility.reveal\&hl=en\_US\&gl=US}}.
\newblock
\newblock
\shownote{(Accessed on 03/30/2024).}


\bibitem[\protect\citeauthoryear{Guerreiro and Gon{\c{c}}alves}{Guerreiro and Gon{\c{c}}alves}{2014}]%
        {guerreiro2014text}
\bibfield{author}{\bibinfo{person}{Jo{\~a}o Guerreiro} {and} \bibinfo{person}{Daniel Gon{\c{c}}alves}.} \bibinfo{year}{2014}\natexlab{}.
\newblock \showarticletitle{Text-to-speeches: evaluating the perception of concurrent speech by blind people}. In \bibinfo{booktitle}{\emph{Proceedings of the 16th international ACM SIGACCESS conference on Computers \& accessibility}}. \bibinfo{pages}{169--176}.
\newblock


\bibitem[\protect\citeauthoryear{Guo, Chen, Qi, White, Ghosh, Asakawa, and Bigham}{Guo et~al\mbox{.}}{2016}]%
        {guo2016vizlens}
\bibfield{author}{\bibinfo{person}{Anhong Guo}, \bibinfo{person}{Xiang'Anthony' Chen}, \bibinfo{person}{Haoran Qi}, \bibinfo{person}{Samuel White}, \bibinfo{person}{Suman Ghosh}, \bibinfo{person}{Chieko Asakawa}, {and} \bibinfo{person}{Jeffrey~P Bigham}.} \bibinfo{year}{2016}\natexlab{}.
\newblock \showarticletitle{Vizlens: A robust and interactive screen reader for interfaces in the real world}. In \bibinfo{booktitle}{\emph{Proceedings of the 29th Annual Symposium on User Interface Software and Technology}}. \bibinfo{pages}{651--664}.
\newblock


\bibitem[\protect\citeauthoryear{Guo, Kim, Chen, Yeh, Hudson, Mankoff, and Bigham}{Guo et~al\mbox{.}}{2017}]%
        {guo2017facade}
\bibfield{author}{\bibinfo{person}{Anhong Guo}, \bibinfo{person}{Jeeeun Kim}, \bibinfo{person}{Xiang'Anthony' Chen}, \bibinfo{person}{Tom Yeh}, \bibinfo{person}{Scott~E Hudson}, \bibinfo{person}{Jennifer Mankoff}, {and} \bibinfo{person}{Jeffrey~P Bigham}.} \bibinfo{year}{2017}\natexlab{}.
\newblock \showarticletitle{Facade: Auto-generating tactile interfaces to appliances}. In \bibinfo{booktitle}{\emph{Proceedings of the 2017 CHI Conference on Human Factors in Computing Systems}}. \bibinfo{pages}{5826--5838}.
\newblock


\bibitem[\protect\citeauthoryear{Harada, Sato, Adams, Kurniawan, Takagi, and Asakawa}{Harada et~al\mbox{.}}{2013}]%
        {harada2013accessible}
\bibfield{author}{\bibinfo{person}{Susumu Harada}, \bibinfo{person}{Daisuke Sato}, \bibinfo{person}{Dustin~W Adams}, \bibinfo{person}{Sri Kurniawan}, \bibinfo{person}{Hironobu Takagi}, {and} \bibinfo{person}{Chieko Asakawa}.} \bibinfo{year}{2013}\natexlab{}.
\newblock \showarticletitle{Accessible photo album: enhancing the photo sharing experience for people with visual impairment}. In \bibinfo{booktitle}{\emph{Proceedings of the SIGCHI conference on human factors in computing systems}}. \bibinfo{pages}{2127--2136}.
\newblock


\bibitem[\protect\citeauthoryear{Hartson and Pyla}{Hartson and Pyla}{2012}]%
        {hartson2012ux}
\bibfield{author}{\bibinfo{person}{Rex Hartson} {and} \bibinfo{person}{Pardha~S Pyla}.} \bibinfo{year}{2012}\natexlab{}.
\newblock \bibinfo{booktitle}{\emph{The UX Book: Process and guidelines for ensuring a quality user experience}}.
\newblock \bibinfo{publisher}{Elsevier}.
\newblock


\bibitem[\protect\citeauthoryear{He, Wan, Findlater, and Froehlich}{He et~al\mbox{.}}{2017}]%
        {he2017tactile}
\bibfield{author}{\bibinfo{person}{Liang He}, \bibinfo{person}{Zijian Wan}, \bibinfo{person}{Leah Findlater}, {and} \bibinfo{person}{Jon~E Froehlich}.} \bibinfo{year}{2017}\natexlab{}.
\newblock \showarticletitle{TacTILE: a preliminary toolchain for creating accessible graphics with 3D-printed overlays and auditory annotations}. In \bibinfo{booktitle}{\emph{Proceedings of the 19th International ACM SIGACCESS Conference on Computers and Accessibility}}. \bibinfo{pages}{397--398}.
\newblock


\bibitem[\protect\citeauthoryear{Inc.}{Inc.}{2021}]%
        {Accessib51:online}
\bibfield{author}{\bibinfo{person}{Apple Inc.}} \bibinfo{year}{2021}\natexlab{}.
\newblock \bibinfo{title}{Accessibility - Vision - Apple (CA)}.
\newblock \bibinfo{howpublished}{\url{https://www.apple.com/ca/accessibility/vision/}}.
\newblock
\newblock
\shownote{(Accessed on 03/16/2021).}


\bibitem[\protect\citeauthoryear{Inc.}{Inc.}{2012}]%
        {VideoCon42:online}
\bibfield{author}{\bibinfo{person}{Zoom Video~Communications Inc.}} \bibinfo{year}{2012}\natexlab{}.
\newblock \bibinfo{title}{Video Conferencing, Cloud Phone, Webinars, Chat, Virtual Events | Zoom}.
\newblock \bibinfo{howpublished}{\url{https://zoom.us/}}.
\newblock
\newblock
\shownote{(Accessed on 07/20/2021).}


\bibitem[\protect\citeauthoryear{Jones, Bartlett, and Cooke}{Jones et~al\mbox{.}}{2019}]%
        {jones2019analysis}
\bibfield{author}{\bibinfo{person}{Nabila Jones}, \bibinfo{person}{Hannah~Elizabeth Bartlett}, {and} \bibinfo{person}{Richard Cooke}.} \bibinfo{year}{2019}\natexlab{}.
\newblock \showarticletitle{An analysis of the impact of visual impairment on activities of daily living and vision-related quality of life in a visually impaired adult population}.
\newblock \bibinfo{journal}{\emph{British Journal of Visual Impairment}} \bibinfo{volume}{37}, \bibinfo{number}{1} (\bibinfo{year}{2019}), \bibinfo{pages}{50--63}.
\newblock


\bibitem[\protect\citeauthoryear{Kane, Bigham, and Wobbrock}{Kane et~al\mbox{.}}{2008}]%
        {kane2008slide}
\bibfield{author}{\bibinfo{person}{Shaun~K Kane}, \bibinfo{person}{Jeffrey~P Bigham}, {and} \bibinfo{person}{Jacob~O Wobbrock}.} \bibinfo{year}{2008}\natexlab{}.
\newblock \showarticletitle{Slide rule: making mobile touch screens accessible to blind people using multi-touch interaction techniques}. In \bibinfo{booktitle}{\emph{Proceedings of the 10th international ACM SIGACCESS conference on Computers and accessibility}}. \bibinfo{pages}{73--80}.
\newblock


\bibitem[\protect\citeauthoryear{Kane, Wobbrock, and Ladner}{Kane et~al\mbox{.}}{2011}]%
        {kane2011usable}
\bibfield{author}{\bibinfo{person}{Shaun~K Kane}, \bibinfo{person}{Jacob~O Wobbrock}, {and} \bibinfo{person}{Richard~E Ladner}.} \bibinfo{year}{2011}\natexlab{}.
\newblock \showarticletitle{Usable gestures for blind people: understanding preference and performance}. In \bibinfo{booktitle}{\emph{Proceedings of the SIGCHI Conference on Human Factors in Computing Systems}}. \bibinfo{pages}{413--422}.
\newblock


\bibitem[\protect\citeauthoryear{Kasneci, Se{\ss}ler, K{\"u}chemann, Bannert, Dementieva, Fischer, Gasser, Groh, G{\"u}nnemann, H{\"u}llermeier, et~al\mbox{.}}{Kasneci et~al\mbox{.}}{2023}]%
        {kasneci2023chatgpt}
\bibfield{author}{\bibinfo{person}{Enkelejda Kasneci}, \bibinfo{person}{Kathrin Se{\ss}ler}, \bibinfo{person}{Stefan K{\"u}chemann}, \bibinfo{person}{Maria Bannert}, \bibinfo{person}{Daryna Dementieva}, \bibinfo{person}{Frank Fischer}, \bibinfo{person}{Urs Gasser}, \bibinfo{person}{Georg Groh}, \bibinfo{person}{Stephan G{\"u}nnemann}, \bibinfo{person}{Eyke H{\"u}llermeier}, {et~al\mbox{.}}} \bibinfo{year}{2023}\natexlab{}.
\newblock \showarticletitle{ChatGPT for good? On opportunities and challenges of large language models for education}.
\newblock \bibinfo{journal}{\emph{Learning and individual differences}}  \bibinfo{volume}{103} (\bibinfo{year}{2023}), \bibinfo{pages}{102274}.
\newblock


\bibitem[\protect\citeauthoryear{Khusid and Shardin}{Khusid and Shardin}{2011}]%
        {AnOnline70:online}
\bibfield{author}{\bibinfo{person}{Andrey Khusid} {and} \bibinfo{person}{Oleg Shardin}.} \bibinfo{year}{2011}\natexlab{}.
\newblock \bibinfo{title}{An Online Visual Collaboration Platform for Teamwork | Miro}.
\newblock \bibinfo{howpublished}{\url{https://miro.com/}}.
\newblock
\newblock
\shownote{(Accessed on 06/02/2021).}


\bibitem[\protect\citeauthoryear{Kianpisheh, Li, and Truong}{Kianpisheh et~al\mbox{.}}{2019}]%
        {kianpisheh2019face}
\bibfield{author}{\bibinfo{person}{Mohammad Kianpisheh}, \bibinfo{person}{Franklin~Mingzhe Li}, {and} \bibinfo{person}{Khai~N Truong}.} \bibinfo{year}{2019}\natexlab{}.
\newblock \showarticletitle{Face recognition assistant for people with visual impairments}.
\newblock \bibinfo{journal}{\emph{Proceedings of the ACM on Interactive, Mobile, Wearable and Ubiquitous Technologies}} \bibinfo{volume}{3}, \bibinfo{number}{3} (\bibinfo{year}{2019}), \bibinfo{pages}{1--24}.
\newblock


\bibitem[\protect\citeauthoryear{Kostyra, {\.Z}akowska-Biemans, {\'S}niegocka, and Piotrowska}{Kostyra et~al\mbox{.}}{2017}]%
        {kostyra2017food}
\bibfield{author}{\bibinfo{person}{Eliza Kostyra}, \bibinfo{person}{Sylwia {\.Z}akowska-Biemans}, \bibinfo{person}{Katarzyna {\'S}niegocka}, {and} \bibinfo{person}{Anna Piotrowska}.} \bibinfo{year}{2017}\natexlab{}.
\newblock \showarticletitle{Food shopping, sensory determinants of food choice and meal preparation by visually impaired people. Obstacles and expectations in daily food experiences}.
\newblock \bibinfo{journal}{\emph{Appetite}}  \bibinfo{volume}{113} (\bibinfo{year}{2017}), \bibinfo{pages}{14--22}.
\newblock


\bibitem[\protect\citeauthoryear{Kroger}{Kroger}{2023}]%
        {KrogerCh9:online}
\bibfield{author}{\bibinfo{person}{Kroger}.} \bibinfo{year}{2023}\natexlab{}.
\newblock \bibinfo{title}{Kroger Chefbot – The Kroger Co.}
\newblock \bibinfo{howpublished}{\url{https://www.thekrogerco.com/kroger-chefbot/}}.
\newblock
\newblock
\shownote{(Accessed on 05/11/2023).}


\bibitem[\protect\citeauthoryear{Kuoppam{\"a}ki, Tuncer, Eriksson, and McMillan}{Kuoppam{\"a}ki et~al\mbox{.}}{2021}]%
        {kuoppamaki2021designing}
\bibfield{author}{\bibinfo{person}{Sanna Kuoppam{\"a}ki}, \bibinfo{person}{Sylvaine Tuncer}, \bibinfo{person}{Sara Eriksson}, {and} \bibinfo{person}{Donald McMillan}.} \bibinfo{year}{2021}\natexlab{}.
\newblock \showarticletitle{Designing Kitchen Technologies for Ageing in Place: A Video Study of Older Adults' Cooking at Home}.
\newblock \bibinfo{journal}{\emph{Proceedings of the ACM on interactive, mobile, wearable and ubiquitous technologies}} \bibinfo{volume}{5}, \bibinfo{number}{2} (\bibinfo{year}{2021}), \bibinfo{pages}{1--19}.
\newblock


\bibitem[\protect\citeauthoryear{Kusu, Makino, Shioi, and Hatano}{Kusu et~al\mbox{.}}{2017}]%
        {kusu2017calculating}
\bibfield{author}{\bibinfo{person}{Kazuma Kusu}, \bibinfo{person}{Nozomi Makino}, \bibinfo{person}{Takamitsu Shioi}, {and} \bibinfo{person}{Kenji Hatano}.} \bibinfo{year}{2017}\natexlab{}.
\newblock \showarticletitle{Calculating Cooking Recipe's Difficulty based on Cooking Activities}. In \bibinfo{booktitle}{\emph{Proceedings of the 9th Workshop on Multimedia for Cooking and Eating Activities in conjunction with The 2017 International Joint Conference on Artificial Intelligence}}. \bibinfo{pages}{19--24}.
\newblock


\bibitem[\protect\citeauthoryear{Lazar, Allen, Kleinman, and Malarkey}{Lazar et~al\mbox{.}}{2007}]%
        {lazar2007frustrates}
\bibfield{author}{\bibinfo{person}{Jonathan Lazar}, \bibinfo{person}{Aaron Allen}, \bibinfo{person}{Jason Kleinman}, {and} \bibinfo{person}{Chris Malarkey}.} \bibinfo{year}{2007}\natexlab{}.
\newblock \showarticletitle{What frustrates screen reader users on the web: A study of 100 blind users}.
\newblock \bibinfo{journal}{\emph{International Journal of human-computer interaction}} \bibinfo{volume}{22}, \bibinfo{number}{3} (\bibinfo{year}{2007}), \bibinfo{pages}{247--269}.
\newblock


\bibitem[\protect\citeauthoryear{Lei, Fan, and Thang}{Lei et~al\mbox{.}}{2022}]%
        {lei2022shake}
\bibfield{author}{\bibinfo{person}{Wentao Lei}, \bibinfo{person}{Mingming Fan}, {and} \bibinfo{person}{Juliann Thang}.} \bibinfo{year}{2022}\natexlab{}.
\newblock \showarticletitle{“I Shake The Package To Check If It’s Mine” A Study of Package Fetching Practices and Challenges of Blind and Low Vision People in China}. In \bibinfo{booktitle}{\emph{Proceedings of the 2022 CHI Conference on Human Factors in Computing Systems}}. \bibinfo{pages}{1--15}.
\newblock


\bibitem[\protect\citeauthoryear{Leporini, Buzzi, and Buzzi}{Leporini et~al\mbox{.}}{2012}]%
        {leporini2012interacting}
\bibfield{author}{\bibinfo{person}{Barbara Leporini}, \bibinfo{person}{Maria~Claudia Buzzi}, {and} \bibinfo{person}{Marina Buzzi}.} \bibinfo{year}{2012}\natexlab{}.
\newblock \showarticletitle{Interacting with mobile devices via VoiceOver: usability and accessibility issues}. In \bibinfo{booktitle}{\emph{Proceedings of the 24th Australian Computer-Human Interaction Conference}}. \bibinfo{pages}{339--348}.
\newblock


\bibitem[\protect\citeauthoryear{Li, Chen, Fan, and Truong}{Li et~al\mbox{.}}{2019}]%
        {li2019fmt}
\bibfield{author}{\bibinfo{person}{Franklin~Mingzhe Li}, \bibinfo{person}{Di~Laura Chen}, \bibinfo{person}{Mingming Fan}, {and} \bibinfo{person}{Khai~N Truong}.} \bibinfo{year}{2019}\natexlab{}.
\newblock \showarticletitle{FMT: A Wearable Camera-Based Object Tracking Memory Aid for Older Adults}.
\newblock \bibinfo{journal}{\emph{Proceedings of the ACM on Interactive, Mobile, Wearable and Ubiquitous Technologies}} \bibinfo{volume}{3}, \bibinfo{number}{3} (\bibinfo{year}{2019}), \bibinfo{pages}{1--25}.
\newblock


\bibitem[\protect\citeauthoryear{Li, Chen, Fan, and Truong}{Li et~al\mbox{.}}{2021a}]%
        {li2021choose}
\bibfield{author}{\bibinfo{person}{Franklin~Mingzhe Li}, \bibinfo{person}{Di~Laura Chen}, \bibinfo{person}{Mingming Fan}, {and} \bibinfo{person}{Khai~N Truong}.} \bibinfo{year}{2021}\natexlab{a}.
\newblock \showarticletitle{“I Choose Assistive Devices That Save My Face” A Study on Perceptions of Accessibility and Assistive Technology Use Conducted in China}. In \bibinfo{booktitle}{\emph{Proceedings of the 2021 CHI Conference on Human Factors in Computing Systems}}. \bibinfo{pages}{1--14}.
\newblock


\bibitem[\protect\citeauthoryear{Li, Dorst, Cederberg, and Carrington}{Li et~al\mbox{.}}{2021b}]%
        {li2021non}
\bibfield{author}{\bibinfo{person}{Franklin~Mingzhe Li}, \bibinfo{person}{Jamie Dorst}, \bibinfo{person}{Peter Cederberg}, {and} \bibinfo{person}{Patrick Carrington}.} \bibinfo{year}{2021}\natexlab{b}.
\newblock \showarticletitle{Non-visual cooking: exploring practices and challenges of meal preparation by people with visual impairments}. In \bibinfo{booktitle}{\emph{Proceedings of the 23rd International ACM SIGACCESS Conference on Computers and Accessibility}}. \bibinfo{pages}{1--11}.
\newblock


\bibitem[\protect\citeauthoryear{Li, Liu, Kane, and Carrington}{Li et~al\mbox{.}}{2024}]%
        {li2024contextual}
\bibfield{author}{\bibinfo{person}{Franklin~Mingzhe Li}, \bibinfo{person}{Michael~Xieyang Liu}, \bibinfo{person}{Shaun~K Kane}, {and} \bibinfo{person}{Patrick Carrington}.} \bibinfo{year}{2024}\natexlab{}.
\newblock \showarticletitle{A Contextual Inquiry of People with Vision Impairments in Cooking}. In \bibinfo{booktitle}{\emph{Proceedings of the CHI Conference on Human Factors in Computing Systems}}. \bibinfo{pages}{1--14}.
\newblock


\bibitem[\protect\citeauthoryear{Li, Liu, Zhang, and Carrington}{Li et~al\mbox{.}}{2022a}]%
        {li2022freedom}
\bibfield{author}{\bibinfo{person}{Franklin~Mingzhe Li}, \bibinfo{person}{Michael~Xieyang Liu}, \bibinfo{person}{Yang Zhang}, {and} \bibinfo{person}{Patrick Carrington}.} \bibinfo{year}{2022}\natexlab{a}.
\newblock \showarticletitle{Freedom to Choose: Understanding Input Modality Preferences of People with Upper-body Motor Impairments for Activities of Daily Living}. In \bibinfo{booktitle}{\emph{Proceedings of the 24th International ACM SIGACCESS Conference on Computers and Accessibility}}. \bibinfo{pages}{1--16}.
\newblock


\bibitem[\protect\citeauthoryear{Li, Spektor, Xia, Huh, Cederberg, Gong, Shinohara, and Carrington}{Li et~al\mbox{.}}{2022b}]%
        {li2022feels}
\bibfield{author}{\bibinfo{person}{Franklin~Mingzhe Li}, \bibinfo{person}{Franchesca Spektor}, \bibinfo{person}{Meng Xia}, \bibinfo{person}{Mina Huh}, \bibinfo{person}{Peter Cederberg}, \bibinfo{person}{Yuqi Gong}, \bibinfo{person}{Kristen Shinohara}, {and} \bibinfo{person}{Patrick Carrington}.} \bibinfo{year}{2022}\natexlab{b}.
\newblock \showarticletitle{“It Feels Like Taking a Gamble”: Exploring Perceptions, Practices, and Challenges of Using Makeup and Cosmetics for People with Visual Impairments}. In \bibinfo{booktitle}{\emph{CHI Conference on Human Factors in Computing Systems}}. \bibinfo{pages}{1--15}.
\newblock


\bibitem[\protect\citeauthoryear{Li, Fan, and Truong}{Li et~al\mbox{.}}{2017}]%
        {li2017braillesketch}
\bibfield{author}{\bibinfo{person}{Mingzhe Li}, \bibinfo{person}{Mingming Fan}, {and} \bibinfo{person}{Khai~N Truong}.} \bibinfo{year}{2017}\natexlab{}.
\newblock \showarticletitle{BrailleSketch: A gesture-based text input method for people with visual impairments}. In \bibinfo{booktitle}{\emph{Proceedings of the 19th International ACM SIGACCESS Conference on Computers and Accessibility}}. \bibinfo{pages}{12--21}.
\newblock


\bibitem[\protect\citeauthoryear{Lin and Han}{Lin and Han}{2014}]%
        {lin2014context}
\bibfield{author}{\bibinfo{person}{Qing Lin} {and} \bibinfo{person}{Youngjoon Han}.} \bibinfo{year}{2014}\natexlab{}.
\newblock \showarticletitle{A context-aware-based audio guidance system for blind people using a multimodal profile model}.
\newblock \bibinfo{journal}{\emph{Sensors}} \bibinfo{volume}{14}, \bibinfo{number}{10} (\bibinfo{year}{2014}), \bibinfo{pages}{18670--18700}.
\newblock


\bibitem[\protect\citeauthoryear{Liu, Wu, Chen, Li, Kittur, and Myers}{Liu et~al\mbox{.}}{2024}]%
        {liu2024selenite}
\bibfield{author}{\bibinfo{person}{Michael~Xieyang Liu}, \bibinfo{person}{Tongshuang Wu}, \bibinfo{person}{Tianying Chen}, \bibinfo{person}{Franklin~Mingzhe Li}, \bibinfo{person}{Aniket Kittur}, {and} \bibinfo{person}{Brad~A Myers}.} \bibinfo{year}{2024}\natexlab{}.
\newblock \showarticletitle{Selenite: Scaffolding Online Sensemaking with Comprehensive Overviews Elicited from Large Language Models}. In \bibinfo{booktitle}{\emph{Proceedings of the CHI Conference on Human Factors in Computing Systems}}. \bibinfo{pages}{1--26}.
\newblock


\bibitem[\protect\citeauthoryear{Liu, Carrington, Chen, and Pavel}{Liu et~al\mbox{.}}{2021}]%
        {liu2021makes}
\bibfield{author}{\bibinfo{person}{Xingyu Liu}, \bibinfo{person}{Patrick Carrington}, \bibinfo{person}{Xiang'Anthony' Chen}, {and} \bibinfo{person}{Amy Pavel}.} \bibinfo{year}{2021}\natexlab{}.
\newblock \showarticletitle{What makes videos accessible to blind and visually impaired people?}. In \bibinfo{booktitle}{\emph{Proceedings of the 2021 CHI Conference on Human Factors in Computing Systems}}. \bibinfo{pages}{1--14}.
\newblock


\bibitem[\protect\citeauthoryear{LLC}{LLC}{2021}]%
        {Getstart6:online}
\bibfield{author}{\bibinfo{person}{Google LLC}.} \bibinfo{year}{2021}\natexlab{}.
\newblock \bibinfo{title}{Get started on Android with TalkBack - Android Accessibility Help}.
\newblock \bibinfo{howpublished}{\url{https://support.google.com/accessibility/android/answer/6283677?hl=en}}.
\newblock
\newblock
\shownote{(Accessed on 03/16/2021).}


\bibitem[\protect\citeauthoryear{Microsoft}{Microsoft}{2023}]%
        {SeeingAI86:online}
\bibfield{author}{\bibinfo{person}{Microsoft}.} \bibinfo{year}{2023}\natexlab{}.
\newblock \bibinfo{title}{Seeing AI App from Microsoft}.
\newblock \bibinfo{howpublished}{\url{https://www.microsoft.com/en-us/ai/seeing-ai}}.
\newblock
\newblock
\shownote{(Accessed on 09/03/2023).}


\bibitem[\protect\citeauthoryear{Min, Jiang, Liu, Rui, and Jain}{Min et~al\mbox{.}}{2019}]%
        {min2019survey}
\bibfield{author}{\bibinfo{person}{Weiqing Min}, \bibinfo{person}{Shuqiang Jiang}, \bibinfo{person}{Linhu Liu}, \bibinfo{person}{Yong Rui}, {and} \bibinfo{person}{Ramesh Jain}.} \bibinfo{year}{2019}\natexlab{}.
\newblock \showarticletitle{A survey on food computing}.
\newblock \bibinfo{journal}{\emph{ACM Computing Surveys (CSUR)}} \bibinfo{volume}{52}, \bibinfo{number}{5} (\bibinfo{year}{2019}), \bibinfo{pages}{1--36}.
\newblock


\bibitem[\protect\citeauthoryear{Morris, Zolyomi, Yao, Bahram, Bigham, and Kane}{Morris et~al\mbox{.}}{2016}]%
        {morris2016most}
\bibfield{author}{\bibinfo{person}{Meredith~Ringel Morris}, \bibinfo{person}{Annuska Zolyomi}, \bibinfo{person}{Catherine Yao}, \bibinfo{person}{Sina Bahram}, \bibinfo{person}{Jeffrey~P Bigham}, {and} \bibinfo{person}{Shaun~K Kane}.} \bibinfo{year}{2016}\natexlab{}.
\newblock \showarticletitle{" With most of it being pictures now, I rarely use it" Understanding Twitter's Evolving Accessibility to Blind Users}. In \bibinfo{booktitle}{\emph{Proceedings of the 2016 CHI conference on human factors in computing systems}}. \bibinfo{pages}{5506--5516}.
\newblock


\bibitem[\protect\citeauthoryear{Morris, Blenkhorn, Crossey, Ngo, Ross, Werner, and Wong}{Morris et~al\mbox{.}}{2006}]%
        {morris2006clearspeech}
\bibfield{author}{\bibinfo{person}{Tim Morris}, \bibinfo{person}{Paul Blenkhorn}, \bibinfo{person}{Luke Crossey}, \bibinfo{person}{Quang Ngo}, \bibinfo{person}{Martin Ross}, \bibinfo{person}{David Werner}, {and} \bibinfo{person}{Christina Wong}.} \bibinfo{year}{2006}\natexlab{}.
\newblock \showarticletitle{Clearspeech: A display reader for the visually handicapped}.
\newblock \bibinfo{journal}{\emph{IEEE Transactions on Neural Systems and Rehabilitation Engineering}} \bibinfo{volume}{14}, \bibinfo{number}{4} (\bibinfo{year}{2006}), \bibinfo{pages}{492--500}.
\newblock


\bibitem[\protect\citeauthoryear{Network}{Network}{2023}]%
        {RecipesD31:online}
\bibfield{author}{\bibinfo{person}{Food Network}.} \bibinfo{year}{2023}\natexlab{}.
\newblock \bibinfo{title}{Recipes, Dinners and Easy Meal Ideas | Food Network}.
\newblock \bibinfo{howpublished}{\url{https://www.foodnetwork.com/recipes}}.
\newblock
\newblock
\shownote{(Accessed on 04/18/2023).}


\bibitem[\protect\citeauthoryear{OpenAI}{OpenAI}{2023}]%
        {Introduc4:online}
\bibfield{author}{\bibinfo{person}{OpenAI}.} \bibinfo{year}{2023}\natexlab{}.
\newblock \bibinfo{title}{Introducing ChatGPT}.
\newblock \bibinfo{howpublished}{\url{https://openai.com/blog/chatgpt}}.
\newblock
\newblock
\shownote{(Accessed on 05/11/2023).}


\bibitem[\protect\citeauthoryear{Phutane, Wright, Castro, Shi, Stern, Lawson, and Azenkot}{Phutane et~al\mbox{.}}{2022}]%
        {phutane2022tactile}
\bibfield{author}{\bibinfo{person}{Mahika Phutane}, \bibinfo{person}{Julie Wright}, \bibinfo{person}{Brenda~Veronica Castro}, \bibinfo{person}{Lei Shi}, \bibinfo{person}{Simone~R Stern}, \bibinfo{person}{Holly~M Lawson}, {and} \bibinfo{person}{Shiri Azenkot}.} \bibinfo{year}{2022}\natexlab{}.
\newblock \showarticletitle{Tactile materials in practice: Understanding the experiences of teachers of the visually impaired}.
\newblock \bibinfo{journal}{\emph{ACM Transactions on Accessible Computing (TACCESS)}} \bibinfo{volume}{15}, \bibinfo{number}{3} (\bibinfo{year}{2022}), \bibinfo{pages}{1--34}.
\newblock


\bibitem[\protect\citeauthoryear{Pini, Hayes, Upton, and Corcoran}{Pini et~al\mbox{.}}{2019}]%
        {pini2019ai}
\bibfield{author}{\bibinfo{person}{Azzurra Pini}, \bibinfo{person}{Jer Hayes}, \bibinfo{person}{Connor Upton}, {and} \bibinfo{person}{Medb Corcoran}.} \bibinfo{year}{2019}\natexlab{}.
\newblock \showarticletitle{AI inspired recipes: Designing computationally creative food combos}. In \bibinfo{booktitle}{\emph{Extended Abstracts of the 2019 CHI Conference on Human Factors in Computing Systems}}. \bibinfo{pages}{1--6}.
\newblock


\bibitem[\protect\citeauthoryear{Ribeiro, Barbosa, Okimoto, and Vieira}{Ribeiro et~al\mbox{.}}{2019}]%
        {ribeiro2019information}
\bibfield{author}{\bibinfo{person}{Gisele Yumi~Arabori Ribeiro}, \bibinfo{person}{Maria Lilian de~Ara{\'u}jo Barbosa}, \bibinfo{person}{Maria L{\'u}cia Leite~Ribeiro Okimoto}, {and} \bibinfo{person}{Rafael~Lima Vieira}.} \bibinfo{year}{2019}\natexlab{}.
\newblock \showarticletitle{Information for Tactile Reading: A Study of Tactile Ergonomics of Packaging for Blind People}. In \bibinfo{booktitle}{\emph{Proceedings of the 20th Congress of the International Ergonomics Association (IEA 2018) Volume VII: Ergonomics in Design, Design for All, Activity Theories for Work Analysis and Design, Affective Design 20}}. Springer, \bibinfo{pages}{1682--1688}.
\newblock


\bibitem[\protect\citeauthoryear{Rodrigues, Montague, Nicolau, and Guerreiro}{Rodrigues et~al\mbox{.}}{2015}]%
        {rodrigues2015getting}
\bibfield{author}{\bibinfo{person}{Andr{\'e} Rodrigues}, \bibinfo{person}{Kyle Montague}, \bibinfo{person}{Hugo Nicolau}, {and} \bibinfo{person}{Tiago Guerreiro}.} \bibinfo{year}{2015}\natexlab{}.
\newblock \showarticletitle{Getting smartphones to talkback: Understanding the smartphone adoption process of blind users}. In \bibinfo{booktitle}{\emph{Proceedings of the 17th international acm sigaccess conference on computers \& accessibility}}. \bibinfo{pages}{23--32}.
\newblock


\bibitem[\protect\citeauthoryear{Russomanno, O’Modhrain, Gillespie, and Rodger}{Russomanno et~al\mbox{.}}{2015}]%
        {russomanno2015refreshing}
\bibfield{author}{\bibinfo{person}{Alexander Russomanno}, \bibinfo{person}{Sile O’Modhrain}, \bibinfo{person}{R~Brent Gillespie}, {and} \bibinfo{person}{Matthew~WM Rodger}.} \bibinfo{year}{2015}\natexlab{}.
\newblock \showarticletitle{Refreshing refreshable braille displays}.
\newblock \bibinfo{journal}{\emph{IEEE transactions on haptics}} \bibinfo{volume}{8}, \bibinfo{number}{3} (\bibinfo{year}{2015}), \bibinfo{pages}{287--297}.
\newblock


\bibitem[\protect\citeauthoryear{S{\'a}nchez, Darin, and Andrade}{S{\'a}nchez et~al\mbox{.}}{2015}]%
        {sanchez2015multimodal}
\bibfield{author}{\bibinfo{person}{Jaime S{\'a}nchez}, \bibinfo{person}{Ticianne Darin}, {and} \bibinfo{person}{Rossana Andrade}.} \bibinfo{year}{2015}\natexlab{}.
\newblock \showarticletitle{Multimodal videogames for the cognition of people who are blind: trends and issues}. In \bibinfo{booktitle}{\emph{Universal Access in Human-Computer Interaction. Access to Learning, Health and Well-Being: 9th International Conference, UAHCI 2015, Held as Part of HCI International 2015, Los Angeles, CA, USA, August 2-7, 2015, Proceedings, Part III 9}}. Springer, \bibinfo{pages}{535--546}.
\newblock


\bibitem[\protect\citeauthoryear{Shimomura, Hvannberg, and Hafsteinsson}{Shimomura et~al\mbox{.}}{2010}]%
        {shimomura2010accessibility}
\bibfield{author}{\bibinfo{person}{Yayoi Shimomura}, \bibinfo{person}{Ebba~Thora Hvannberg}, {and} \bibinfo{person}{Hjalmtyr Hafsteinsson}.} \bibinfo{year}{2010}\natexlab{}.
\newblock \showarticletitle{Accessibility of audio and tactile interfaces for young blind people performing everyday tasks}.
\newblock \bibinfo{journal}{\emph{Universal Access in the Information Society}}  \bibinfo{volume}{9} (\bibinfo{year}{2010}), \bibinfo{pages}{297--310}.
\newblock


\bibitem[\protect\citeauthoryear{Tasse and Smith}{Tasse and Smith}{2008}]%
        {tasse2008sour}
\bibfield{author}{\bibinfo{person}{Dan Tasse} {and} \bibinfo{person}{Noah~A Smith}.} \bibinfo{year}{2008}\natexlab{}.
\newblock \showarticletitle{SOUR CREAM: Toward semantic processing of recipes}.
\newblock \bibinfo{journal}{\emph{Carnegie Mellon University, Pittsburgh, Tech. Rep. CMU-LTI-08-005}} (\bibinfo{year}{2008}).
\newblock


\bibitem[\protect\citeauthoryear{Tekin, Coughlan, and Shen}{Tekin et~al\mbox{.}}{2011}]%
        {tekin2011real}
\bibfield{author}{\bibinfo{person}{Ender Tekin}, \bibinfo{person}{James~M Coughlan}, {and} \bibinfo{person}{Huiying Shen}.} \bibinfo{year}{2011}\natexlab{}.
\newblock \showarticletitle{Real-time detection and reading of LED/LCD displays for visually impaired persons}. In \bibinfo{booktitle}{\emph{2011 IEEE Workshop on Applications of Computer Vision (WACV)}}. IEEE, \bibinfo{pages}{491--496}.
\newblock


\bibitem[\protect\citeauthoryear{Teng, Lin, and Adamic}{Teng et~al\mbox{.}}{2011}]%
        {teng2011reciperecommendation}
\bibfield{author}{\bibinfo{person}{ChunYuen Teng}, \bibinfo{person}{Yu{-}Ru Lin}, {and} \bibinfo{person}{Lada~A. Adamic}.} \bibinfo{year}{2011}\natexlab{}.
\newblock \showarticletitle{Recipe recommendation using ingredient networks}.
\newblock \bibinfo{journal}{\emph{CoRR}}  \bibinfo{volume}{abs/1111.3919} (\bibinfo{year}{2011}).
\newblock
\showeprint[arXiv]{1111.3919}
\urldef\tempurl%
\url{http://arxiv.org/abs/1111.3919}
\showURL{%
\tempurl}


\bibitem[\protect\citeauthoryear{Van~Erp, Reynolds, Maynard, Starke, Ib{\'a}{\~n}ez~Mart{\'\i}n, Andres, Leite, Alvarez~de Toledo, Schmidt~Rivera, Trattner, et~al\mbox{.}}{Van~Erp et~al\mbox{.}}{2021}]%
        {van2021using}
\bibfield{author}{\bibinfo{person}{Marieke Van~Erp}, \bibinfo{person}{Christian Reynolds}, \bibinfo{person}{Diana Maynard}, \bibinfo{person}{Alain Starke}, \bibinfo{person}{Rebeca Ib{\'a}{\~n}ez~Mart{\'\i}n}, \bibinfo{person}{Frederic Andres}, \bibinfo{person}{Maria~CA Leite}, \bibinfo{person}{Damien Alvarez~de Toledo}, \bibinfo{person}{Ximena Schmidt~Rivera}, \bibinfo{person}{Christoph Trattner}, {et~al\mbox{.}}} \bibinfo{year}{2021}\natexlab{}.
\newblock \showarticletitle{Using natural language processing and artificial intelligence to explore the nutrition and sustainability of recipes and food}.
\newblock \bibinfo{journal}{\emph{Frontiers in Artificial Intelligence}}  \bibinfo{volume}{3} (\bibinfo{year}{2021}), \bibinfo{pages}{621577}.
\newblock


\bibitem[\protect\citeauthoryear{Vashistha, Cutrell, Dell, and Anderson}{Vashistha et~al\mbox{.}}{2015}]%
        {vashistha2015social}
\bibfield{author}{\bibinfo{person}{Aditya Vashistha}, \bibinfo{person}{Edward Cutrell}, \bibinfo{person}{Nicola Dell}, {and} \bibinfo{person}{Richard Anderson}.} \bibinfo{year}{2015}\natexlab{}.
\newblock \showarticletitle{Social media platforms for low-income blind people in India}. In \bibinfo{booktitle}{\emph{Proceedings of the 17th international ACM SIGACCESS conference on computers \& accessibility}}. \bibinfo{pages}{259--272}.
\newblock


\bibitem[\protect\citeauthoryear{Wang, Zhou, Nguyen, Mondal, Mutlu, and Zhao}{Wang et~al\mbox{.}}{2023}]%
        {wang2023practices}
\bibfield{author}{\bibinfo{person}{Ru Wang}, \bibinfo{person}{Nihan Zhou}, \bibinfo{person}{Tam Nguyen}, \bibinfo{person}{Sanbrita Mondal}, \bibinfo{person}{Bilge Mutlu}, {and} \bibinfo{person}{Yuhang Zhao}.} \bibinfo{year}{2023}\natexlab{}.
\newblock \showarticletitle{Practices and Barriers of Cooking Training for Blind and Low Vision People}. In \bibinfo{booktitle}{\emph{Proceedings of the 25th International ACM SIGACCESS Conference on Computers and Accessibility}}. \bibinfo{pages}{1--5}.
\newblock


\bibitem[\protect\citeauthoryear{Wang, Liang, Huang, Zhang, Li, and Yu}{Wang et~al\mbox{.}}{2021}]%
        {wang2021toward}
\bibfield{author}{\bibinfo{person}{Yujia Wang}, \bibinfo{person}{Wei Liang}, \bibinfo{person}{Haikun Huang}, \bibinfo{person}{Yongqi Zhang}, \bibinfo{person}{Dingzeyu Li}, {and} \bibinfo{person}{Lap-Fai Yu}.} \bibinfo{year}{2021}\natexlab{}.
\newblock \showarticletitle{Toward automatic audio description generation for accessible videos}. In \bibinfo{booktitle}{\emph{Proceedings of the 2021 CHI Conference on Human Factors in Computing Systems}}. \bibinfo{pages}{1--12}.
\newblock


\bibitem[\protect\citeauthoryear{Wiberg}{Wiberg}{2021}]%
        {BeMyEyes54:online}
\bibfield{author}{\bibinfo{person}{Hans~Jørgen Wiberg}.} \bibinfo{year}{2021}\natexlab{}.
\newblock \bibinfo{title}{Be My Eyes - See the world together}.
\newblock \bibinfo{howpublished}{\url{https://www.bemyeyes.com/}}.
\newblock
\newblock
\shownote{(Accessed on 04/10/2021).}


\bibitem[\protect\citeauthoryear{Williams, Neylan, and Hurst}{Williams et~al\mbox{.}}{2013}]%
        {williams2013preliminary}
\bibfield{author}{\bibinfo{person}{Michele~A Williams}, \bibinfo{person}{Callie Neylan}, {and} \bibinfo{person}{Amy Hurst}.} \bibinfo{year}{2013}\natexlab{}.
\newblock \showarticletitle{Preliminary investigation of the limitations fashion presents to those with vision impairments}.
\newblock \bibinfo{journal}{\emph{Fashion Practice}} \bibinfo{volume}{5}, \bibinfo{number}{1} (\bibinfo{year}{2013}), \bibinfo{pages}{81--105}.
\newblock


\bibitem[\protect\citeauthoryear{Wong and Chen}{Wong and Chen}{2003}]%
        {wong2003new}
\bibfield{author}{\bibinfo{person}{Edward~K Wong} {and} \bibinfo{person}{Minya Chen}.} \bibinfo{year}{2003}\natexlab{}.
\newblock \showarticletitle{A new robust algorithm for video text extraction}.
\newblock \bibinfo{journal}{\emph{Pattern Recognition}} \bibinfo{volume}{36}, \bibinfo{number}{6} (\bibinfo{year}{2003}), \bibinfo{pages}{1397--1406}.
\newblock


\bibitem[\protect\citeauthoryear{YouTube}{YouTube}{2023}]%
        {51YouTub12:online}
\bibfield{author}{\bibinfo{person}{YouTube}.} \bibinfo{year}{2023}\natexlab{}.
\newblock \bibinfo{title}{(51) YouTube}.
\newblock \bibinfo{howpublished}{\url{https://www.youtube.com/}}.
\newblock
\newblock
\shownote{(Accessed on 05/11/2023).}


\bibitem[\protect\citeauthoryear{Zhou, M{\"u}ller, Holzinger, and Chen}{Zhou et~al\mbox{.}}{2023}]%
        {zhou2023ethical}
\bibfield{author}{\bibinfo{person}{Jianlong Zhou}, \bibinfo{person}{Heimo M{\"u}ller}, \bibinfo{person}{Andreas Holzinger}, {and} \bibinfo{person}{Fang Chen}.} \bibinfo{year}{2023}\natexlab{}.
\newblock \showarticletitle{Ethical ChatGPT: Concerns, challenges, and commandments}.
\newblock \bibinfo{journal}{\emph{arXiv preprint arXiv:2305.10646}} (\bibinfo{year}{2023}).
\newblock


\end{thebibliography}


\end{document}